\documentclass[a4paper,11pt]{article}
\pdfoutput=1 

\usepackage{jcappub} 

\usepackage[T1]{fontenc} 

\usepackage{appendix}

\title{Constraints on Reionization from Planck Low-$\ell$ EE Polarization and the Lyman-Series Forests}

\author[a]{Garett Lopez,}
\author[a]{Anson D'Aloisio,}
\author[b,c,d]{Christopher Cain}

\affiliation[a]{Department of Physics and Astronomy, University of California, Riverside,
\\900 University Ave., Riverside, CA, 92521, USA}
\affiliation[b]{School of Earth and Space exploration, Arizona State University, Tempe, AZ 85281, USA}
\affiliation[c]{KICC - Kavli Institute for Cosmology Cambridge, Madingley Road, Cambridge CB3 0HA, UK}
\affiliation[d]{Institute of Astronomy, University of Cambridge, Madingley Road, Cambridge CB3 0HA, UK}

\emailAdd{garett.lopez@ucr.edu}
\emailAdd{ansond@ucr.edu}
\emailAdd{clc218@cam.ac.uk}

\abstract{CMB measurements tightly constrain the Thomson scattering optical depth ($\tau_{\rm es}$), but they still permit a broad range of degenerate reionization histories.  Incorporating Lyman-series forest constraints on the end of reionization can alleviate the degeneracy and narrow the range of allowed histories.  We explore this effect using forest measurements combined with full Planck likelihoods. We apply upper limits on the neutral fraction at $4.9\lesssim z\lesssim6.9$ derived from the dark pixel fraction, as well as a lower limit at $z=5.8$ from damping-wing-like features on the edges of long dark gaps in stacked forest spectra. We find that neither the choice of parametric vs. non-parametric reionization model, nor the choice among different forest data combinations, produces a significant shift in $\tau_{\rm es}$ or the other cosmological parameters. Instead, forest constraints primarily redistribute $\tau_{\rm es}$ in redshift. Combining dark pixel, damping-wing-like, and CMB measurements, we find $z_{50}=7.7^{+0.7}_{-0.8}$ and a 95\% upper limit of $\Delta z_{90}<11.4$ for the midpoint and duration of reionization, respectively. The asymmetry of the reionization history ($A_z$) is not well constrained by these data.  Stronger constraints can be obtained using Ly$\alpha$ forest measurements of the mean IGM photoionization rate, at the cost of requiring more detailed modeling. Using a new semi-analytic reionization model, we obtain (model-dependent) constraints of $z_{50} = 7.4^{+0.2}_{-0.3}$, $\Delta z_{90}=6.0^{+0.8}_{-1.0}$, and $A_z = 2.4^{+0.3}_{-0.7}$. We also find evidence that the ionizing emissivity of reionization sources declines after $z=7-8$, perhaps signaling a decline in the escape fraction and/or the onset of a feedback process. Our results motivate the development of forest likelihood products and fast forward models for its transmission statistics.}


\begin{document}

\maketitle
\flushbottom

\section{Introduction}
\label{sec:intro}

Cosmic reionization marks the transition of the intergalactic medium (IGM) from a predominantly neutral state into the highly ionized medium observed at later times. The reionization history encodes information about the formation and evolution of the first ionizing sources, the escape of ionizing radiation into the IGM, and the density structure of the Universe. Extracting this information, however, has proven extremely challenging; it requires combining complementary constraints from a broad range of observables.

The large-scale Cosmic Microwave Background (CMB) $E$-mode polarization signal provides an important integral constraint on the reionization history. Thomson scattering by free electrons generates the ``reionization bump'' in the low-$\ell$ EE power spectrum \cite{1996ApJ...456..413N, 1997PhRvD..55.1822Z}, while suppressing the primary temperature and polarization anisotropies at smaller angular scales by $e^{-2\tau_{\rm es}}$, where $\tau_{\rm es}$ is the reionization Thomson scattering optical depth. In many CMB analyses, reionization is represented by a rapid and symmetric transition characterized by a single free parameter: the midpoint of reionization. Within this restricted family of histories, $\tau_{\rm es}$ maps almost uniquely onto the reionization midpoint and is therefore commonly described as primarily a constraint on that midpoint.

Recent analyses have emphasized, however, that the full low-$\ell$ polarization spectrum contains information beyond a single parameter. This information may be extracted using physics-based reionization models, flexible parametric histories, redshift bins, or a principal-component framework \cite{2020PhRvL.125g1301H, Qin2020-kf, 2021MNRAS.507.2405C, Heinrich2021-ik, 2021PhRvD.104l3549P, Wu2021-od, 2025PhRvD.111d3532P, Ilic2025-xj}. Current Planck data robustly constrain $\tau_{\rm es}$ across a range of such modeling approaches, but the inferred reionization history remains considerably more sensitive to the adopted model and its implicit priors \citep{Ilic2025-xj}. Joint analyses
with astronomical observations have nevertheless begun to place constraints on additional parameters such as the duration and asymmetry of reionization \cite[e.g.][]{2020PhRvL.125g1301H,2021PhRvD.104l3549P,
2025PhRvD.111d3532P}. These results motivate constraining reionization models directly with CMB likelihoods, rather than compressing the CMB information into a $\tau_{\rm es}$ constraint under a fixed reionization prescription.

Meanwhile, substantial progress has been made in constraining the last phases of reionization using the Lyman-series forests of high-redshift quasars. The rapidly evolving mean transmission, dark gap statistics, and large scatter in the Ly$\alpha$ effective optical depth together suggest that the IGM was undergoing a transition at $5\lesssim z\lesssim6$  \citep{becker15, 2018arXiv180704229E, Bosman2021,2021ApJ...923..223Z,2022ApJ...932...76Z,
2023MNRAS.525.4093G}. Additional evidence comes from the presence of damping-wing-like features in forest spectra \cite{2024arXiv240508885B, Zhu2024-aq, 2024arXiv240512273S}, as well as the rapid evolution in the mean free path of ionizing photons during this epoch \cite{2021MNRAS.508.1853B, 2023ApJ...955..115Z, 2024MNRAS.533..676S, 2024ApJ...965..134D}. Forward modeling of these observations favors a late completion of reionization, with a residual tail extending below $z=6$ \citep{2019MNRAS.485L..24K,  2020MNRAS.494.3080N, 2019arXiv190512640K, Cain2021-yt, 2023MNRAS.525.4093G, 2023ApJ...955..115Z, 2024MNRAS.530.5209R, 
2025PASA...42...49Q} Thus, the forest should no longer be regarded simply as requiring that reionization be complete by approximately $z=6$. Instead, it provides quantitative information about the final $\sim10$--$20\%$ of the transition.

The complementarity between the CMB and the Lyman-series forests has important implications for the interpretation of $\tau_{\rm es}$ measurements. Ref. \cite{2025arXiv250515899C} pointed out that once the endpoint of reionization is constrained to the relatively narrow interval preferred by forest data, $\tau_{\rm es}$ becomes sensitive to the duration of reionization as well as its midpoint. For an unconstrained endpoint, an increase in $\tau_{\rm es}$ can generally be produced by shifting a short reionization history to higher redshift. If the tail end of reionization is instead restricted by the forests, an earlier midpoint generally implies a more extended history and therefore a longer duration. In this sense, the forest changes the degeneracies and physical information carried by the CMB optical depth. 

The analysis of \cite{2025arXiv250515899C} demonstrated this effect by examining the dependence of $\tau_{\rm es}$ on the midpoint, duration, and asymmetry of parametric reionization histories subject to an approximate forest constraint on the endpoint. It did not, however, perform a joint Bayesian analysis of the reionization and cosmological parameters using the full CMB temperature and polarization likelihoods. Such an analysis is important for at least two reasons. First, different reionization histories produce different low-$\ell$ EE spectra even when they have similar values of $\tau_{\rm es}$. Second, the high-$\ell$ CMB spectra constrain the combination $A_{\rm s}e^{-2\tau_{\rm es}}$, where $A_s$ is the amplitude of primordial scalar fluctuations, so that the reionization parameters must be inferred jointly with $A_s$ and the other cosmological parameters. 

In this work, we carry out a joint inference using the full Planck
temperature and polarization likelihoods. Our central goal is to determine how the different Lyman-series forest constraints alter reionization posteriors inferred from the CMB. To this end, we explore the use of different reionization models including the parametric model of \cite{Trac2021-xm} (used in \cite{2025arXiv250515899C}) and the non-parametric \textsc{FlexKnot} approach \cite{Millea2018-xl}.  We will also apply the recently developed physical model of Ref. \cite{Cain2026-op} which connects the ionizing source emissivity to the global ionized fraction and IGM photoionization rate self-consistently.  This model was developed specifically for inference pipelines utilizing Lyman-series forest measurements. Here we will apply it for the first time.   Several distinct forest observables can be used to extract information on the reionization history
and they differ substantially in both constraining power and model
dependence. For instance, the fraction of dark pixels in the Lyman-series forests provide a conservative and nearly model-independent upper limit on the volume-averaged neutral fraction \citep{McGreer2011-zm, Jin2023-bh,Davies2025-lf}.  On the other hand, a conservative but simulation-calibrated lower-limit on the neutral fraction at $z\approx 5.8$ has been derived from Ly$\alpha$ damping-wing-like features on the boundaries of stacked forest dark gaps \cite{Zhu2024-aq}. Among the most constraining statistics is the redshift-evolution of the mean forest transmission \cite{2018arXiv180704229E, Bosman2021}.  Applying it, however, requires detailed forward-modeling.  Another goal of this paper is to explore the information content within this range of forest statistics, and how different combinations alter the joint posteriors. Our
analysis illustrates how information about reionization's end complements the CMB polarization and temperature in constraining the reionization history. 

This paper is organized as follows.  In \S \ref{sec:methods}, we describe our Bayesian inference method, data sets, and likelihoods. In \S \ref{sec:results} we present the results of our joint inferences using CMB, dark pixel, and damping-wing-like forest measurements.  We present posteriors for both parametric and non-parametric models.  In \S \ref{sec:results_Gamma} we discuss how these constraints can be tightened considerably by implementing measurements of the forest transmission fraction.  We conclude in \S \ref{sec:conclusions}. Unless otherwise stated, all distances are reported in comoving units.

\section{Methods}
\label{sec:methods}

\subsection{Bayesian Parameter Inference}
\label{sec:cobaya}

We use the publicly available code, \textit{Cobaya}, to perform Bayesian parameter inference \cite{2021JCAP...05..057T}.  \textit{Cobaya} is widely used for obtaining constraints from cosmological data sets, especially CMB measurements.  Conveniently, it is compatible with Boltzmann theory codes such as Code for Anisotropies in the Microwave Background (CAMB; \cite{2000ApJ...538..473L}) and Cosmic Linear Anisotropy Solving System (CLASS; \cite{2011arXiv1104.2932L}).  It also comes with a number of CMB likelihood functions, e.g. from \textit{Planck}.  For this work, we make two modifications to the \textit{Cobaya} code base.  First, we use a modified version of CAMB that accepts arbitrary reionization histories in the form of a numeric table of redshifts and ionized fraction values, $x_{\rm HII}$.  This affords us flexibility in the range of models that we consider. Second, we construct new likelihoods for the forest constraints used in this work, which we describe below.  

We sample the model parameters using \textit{Cobaya}'s \texttt{mcmc} sampler and pass them to our external forest-likelihood functions.  These functions calculate the reionization history from the parameter values and evaluate the log-likelihood of that history for the forest constraints considered.  In what follows, we explore three different methods for modeling the reionization history: a parametric model (\S \ref{sec:parametric}) in which a functional form is assumed, the more free-form \textsc{FlexKnot} approach of Ref. \cite{Millea2018-xl} (\S \ref{sec:FlexKnot}), and the physical model of Ref. \cite{Cain2026-op} that connects the ionizing emissivity of the source population to the global ionized fraction and photoionization rate of intergalactic hydrogen (\S \ref{sec:results_Gamma}).  The details of these methods will be described in their respective sections.   In addition to the log-likelihood values, our external functions also return $x_{\rm HII}$ evaluated at 50 linearly-spaced points in the range $4<z<30$, which are then passed to CAMB to calculate the CMB power spectra and evaluate the CMB likelihoods.

Unless otherwise noted, we adopt the same uniform priors for cosmological parameters as in Ref. \cite{Tristram2024}: $\theta_{MC} \in [0.0103, 0.0105]$, $\ln(10^{10} A_s) \in [2.6, 3.5]$, $n_s \in [0.9, 1.1]$, $\Omega_bh^2 \in [0.017, 0.027]$, $\Omega_c h^2 \in [0.09, 0.15]$. We also impose a conservative lower limit of $z=4$ for the end of reionization. We use \textit{Cobaya}'s generalized  Gelman-Rubin statistic ($R-1$) to evalulate the convergence of our \texttt{mcmc} chains.  We consider our chains converged when $0.01\lesssim R-1 < 0.03$, where the upper limit of this range represents the most computationally intensive runs (generally the \textsc{FlexKnot} runs). To analyze the samples we use the \texttt{GetDist} code package \cite{Lewis:2019xzd}.

\subsection{Data and Likelihoods}
\label{Data_and_Likelihoods}

\subsubsection{CMB}
\label{sec:CMB_obs}

We combine the following likelihoods in order to leverage the full range of scales measured by \textit{Planck}: 

\begin{itemize}
    \item low-$\ell$ EE: We use the latest version of the Low-$\ell$ Likelihood on Polarised Power-spectra (\texttt{LoLLiPoP}) of Refs. \cite{2021A&A...647A.128T, 2022PhRvD.105h3524T}, based on \textit{{Planck}} PR4 and covering the multipole range $2 \leq \ell \leq 29$.      
    
    \item Low-$\ell$ TT: We use the \textit{Planck} PR3 \texttt{Commander} likelihood in the multipole range $2 \leq \ell \leq 29$ \cite{2020A&A...641A...5P}.
    
    \item high-$\ell$ TTTEEE: We use the nuisance-marginalized \textit{Planck} PR3 \texttt{Plik}$_{\rm lite}$ likelihood over $30\leq \ell \leq 2,508$ for TT, and $30\leq \ell \leq 1,996$ for TE and EE \cite{2020A&A...641A...5P}. We choose this over the full (unmarginalized) \texttt{Plik} likelihood and the more recent \texttt{HiLLiPoP} for computational efficiency, as the small-scale CMB anistropies are relatively less sensitive to the details of the reionization history. 
\end{itemize}

The above combination mixes PR3 and PR4 likelihoods, pairing the PR4 low-$\ell$ EE likelihood with the PR3 high-$\ell$ TTTEEE likelihood (the PR4 analysis retained the PR3 \texttt{Commander} low-$\ell$ TT likelihood, since little improvement in large-scale temperature was expected). In Appendix \ref{sec:PR3_vs_PR4}, we present tests verifying that this choice does not affect our main results.  In summary, we ran an analysis where we replaced \texttt{Plik$_{\rm lite}$} with \texttt{HiLLiPoP} for the high-$\ell$ TTTEEE liklihood and found little difference in the inferred reionization histories. 

\subsubsection{Lyman Series Forest}
\label{sec:Forest_obs}

We consider the combination of CMB measurements with three types of Lyman-series forest measurements: (1) upper limits on the volume-weighted mean neutral fraction, $\langle x_{\rm HI} \rangle_V$, derived from dark pixels in the $z\approx 4.9-6.7$ Ly$\alpha$, Ly$\beta$, and Ly$\gamma$ forests; (2) damping-wing-like features in the $z\approx 5.5-6$ Ly$\alpha$ forest, which place a complementary lower-limit on $\langle x_{\rm HI} \rangle_V$; (3) Ly$\alpha$ forest measurements of the ionizing radiation background strength at $4.9 \leq z \leq 6$, as quantified by the hydrogen photoionization rate, $\Gamma_{\rm HI}$.  The methodology for (3) is more detailed and the results that we obtain from it involve several important caveats.  As such, we defer a description of those measurements to a dedicated section (\S \ref{sec:results_Gamma}).  In this section, we describe the dark pixel and damping-wing-like measurements that form the backbone of our forest likelihoods. 

The dark-pixel constraints on $\langle{x}_{\rm HI} \rangle_V$ exploit the extreme opacity of the high-redshift IGM to Lyman-series absorption. At $z > 5$, a local residual neutral fraction of order $x_{\mathrm{HI}} \sim 10^{-4}$ is enough to render the Ly$\alpha$ forest effectively dark; Ly$\beta$ absorption is also generally strongly saturated at this level. Thus any detected transmission identifies a highly ionized region, whereas a dark region may contain either neutral IGM or ionized gas with a sufficient level of residual neutral hydrogen. The dark-pixel fraction therefore provides a conservative upper limit on the
volume-averaged neutral fraction. While this approach is generally model-independent, in practice one must still make choices on how to process the forest spectra and identify dark pixels.  Following Ref. \cite{McGreer2011-zm}, the forest spectra are typically binned into 3.3 cMpc pixels.\footnote{Ref \cite{McGreer2011-zm} argued any dark gap much smaller than this scale is unlikely to correspond to $x_{\rm HI}$ regions, as the damping wings associated with a Jean-scale neutral region at mean density can be as wide as 3 cMpc.}  Two methods have been employed to identify the dark pixels.  The more conservative ``threshold'' approach simply flags as dark any pixel below some signal-to-noise ratio.  The second approach, termed the negative-pixel method, assumes that truly dark pixels scatter symmetrically above or below the zero point, so the number of dark pixels can be approximated as twice the number of negative pixels.  The negative pixel method provides tighter constraints on $\langle x_{\rm HI} \rangle_{V}$ but is more vulnerable to biasing from imperfect subtraction of the sky signal. In what follows, we use limits obtained from the threshold method. 

We use upper-limits from two sources. First, Ref. \cite{Davies2025-lf} provides the most up-to-date limits at $z=4.85-6.25$ from 34 deep quasar spectra from the extended XQR-30 program \cite{DOdorico2023}.  They report multiple sets of limits by considering different combinations of Ly$\alpha$, Ly$\beta$ and, Ly$\gamma$ forest measurements.  We take the combination that yields their most stringent $\langle x_{\rm HI}\rangle_V$ upper limits at each redshift bin (the bolded values in the 7th column of their Table 2).\footnote{This means the constraints at different redshift bins can be derived from different forest combinations: in particular at $\bar{z}=\{4.875, 5.056 \}$ we take the constraints from Ly$\alpha$ forest alone,  $\bar{z}=5.272$ uses Ly$\alpha$+Ly$\beta$ combination, $\bar{z}=\{ 5.481, 5.654, 5.831, 6.043 \}$ uses the full Ly$\alpha$+Ly$\beta$+Ly$\gamma$ combination, and $\bar{z}=6.225$ considers the Ly$\beta$+Ly$\gamma$ combination.}    To extend the redshift range of the dark pixel constraints, we also include the $z=6.3$, $6.5$, and $6.7$ upper limits from Ref. \cite{Jin2023-bh}, who similarly use dark pixels in the Ly$\alpha$ and Ly$\beta$ forests.  The combination of Refs. \cite{Davies2025-lf, Jin2023-bh} represent the most up-to-date and model-independent upper limits on the volume-weighted mean neutral fraction of the IGM at $4.9 \lesssim z \lesssim 6.7$.  We approximate the likelihood at each redshift bin as one-sided Gaussian distributions.  For a reported $\langle x_{\rm HI}\rangle_V(z_i)$ upper limit at redshift $z_i$, if the predicted neutral fraction, $x_{\rm HI}(z_i) \geq \langle x_{\rm HI} \rangle_V(z_i)$, we approximate the likelihood as a Gaussian centered at $\langle x_{\rm HI} \rangle_V(z_i)$ with standard deviation equal to the reported $1 \sigma$ uncertainty.  On the other hand, if $x_{\rm HI}(z_i) \leq \langle x_{\rm HI}\rangle_V(z_i)$ then the likelihood returns a constant value chosen for continuity at $\langle x_{\rm HI}\rangle_V(z_i)$.  The total likelihood consists of the product of likelihoods at each redshift. 

A complementary lower bound on $\langle x_{\rm HI}\rangle_V$, at $z \simeq 5.8$ has been inferred from a damping-wing-like absorption profile in the stacked Ly$\alpha$ forest adjacent to long Ly$\beta$ dark gaps.\footnote{By definition, a Ly$\beta$ dark gap is a stretch of the forest with no observed transmission in the coeval Ly$\alpha$ and Ly$\beta$ forests.} Following the method originally proposed by Ref. \cite{Malloy2015-ko}, Ref. \cite{Zhu2024-aq} selected 24 Ly$\beta$ dark gaps with lengths $L_{\rm gap} \geq 7\,h^{-1}\,{\rm cMpc}$, aligned the corresponding Ly$\alpha$ forest spectra at the red and blue edges of the gaps, and stacked the transmission (also see \cite{2024arXiv240512273S}). The resulting profile remains suppressed for approximately $1000\,{\rm km\,s^{-1}}$ beyond the gap edges, substantially longer than expected for simulated dark gaps produced entirely by reionized gas, even in the presence of large ionizing background fluctuations. Applying the same stacking and fitting procedure to mock spectra from
radiative transfer simulations, Ref. \cite{Zhu2024-aq} found that only stacks with mean neutral fraction $x_{\rm HI,gap} \gtrsim 0.4$ produced Ly$\alpha$ forest profiles consistent with the observed one within the simulated 95\% interval. Combining this minimum neutral fraction with the measured fraction of the Ly$\alpha$ forest occupied by dark gaps longer than the mean length of their sample, $f_{L \geq 15\,h^{-1}{\rm cMpc}} = 0.152 \pm 0.098$, gives $\langle x_{\rm HI}\rangle_V \gtrsim x_{\rm HI,gap}^{\rm min} f_{L \geq 15\,h^{-1}{\rm cMpc}} = 0.4(0.152 \pm 0.098) = 0.061 \pm 0.039 $ at $z=5.8$. This should be interpreted as a conservative, simulation-calibrated lower bound; it counts only the neutral hydrogen associated with the long-dark-gap population and neglects any neutral gas residing in shorter gaps or elsewhere in the IGM.  Our likelihood for this limit is constructed similarly to the dark pixel constraints, but flipped to reflect it being a lower limit on $\langle x_{\rm HI} \rangle_V$ rather than an upper limit.

In the ensuing sections, we explore four combinations of the observational constraints described above:

\begin{itemize}
    \item \textsc{CMB Only} applies only the CMB likelihoods listed in \ref{sec:CMB_obs}.   
    \item \textsc{DP + CMB} applies the dark pixel upper-limits on $\langle x_{\rm HI}\rangle_V$ from Refs. \cite{Davies2025-lf, Jin2023-bh}, together with the CMB likelihoods.
    \item \textsc{DWlike + CMB} includes the lower limit on $\langle x_{\rm HI}\rangle_V$ derived from the damping-wing-like feature reported by Ref. \cite{Zhu2024-aq} together with the CMB.
    \item \textsc{DP + DWlike + CMB} includes both the upper and lower limits on $\langle x_{\rm HI }\rangle_V$ together with the CMB. 
\end{itemize}

An important caveat is that the dark pixel and damping-wing-like measurements provide limits on the volume-weighted reionization history, whereas the CMB temperature and polarization anisotropies are sensitive to the mass-weighted history.  Mapping these two quantities requires a model for the spatial structure of the reionization process (see e.g. \cite{2026arXiv260518939S}). For simplicity, we assume that the mass-weighted and volume-weighted ionized fractions are approximately the same in our joint CMB-forest analysis. In Appendix \ref{sec:vol_ave}, we use a radiative transfer simulation of reionization to test the effect of this assumption on our results.  We find that correcting for the difference between mass-weighted and volume-weighted has a small effect on our constraints ($\lesssim 5\%$) relative to the uncertainties in our derived reionization histories.

\section{Results}
\label{sec:results}

\subsection{Parametric Reionization Model}
\label{sec:parametric}

We begin by considering a parametric model of the reionization history. We adopt the three-parameter Weibull function previously introduced in Ref. \cite{Trac2021-xm}. This functional form has been used to obtain constraints from the CMB optical depth and patchy kinematic Sunyaev–Zel’dovich effect \citep{2018RNAAS...2..135G, Chen2022, 2025arXiv250515899C}, and to forecast optical depth constraints from 21-cm clustering measurements \citep{2022JCAP...10..007S}. To our knowledge, this is the first analysis in which the Weibull midpoint, duration, and asymmetry parameters are jointly constrained using the full Planck temperature and polarization likelihoods. The reionization history is set by reionization's midpoint ($z_{50}$), duration ($\Delta z_{90}$) and asymmetry ($A_z$).  The midpoint is defined to be the redshift where the mass-weighted ionized fraction of hydrogen reaches 50\%.  The duration and asymmetry parameters are

\begin{equation}
    \label{eqn:Weibull_DzAz}
    \Delta z_{90} = z_{95} - z_{05} \qquad \qquad A_z = \frac{z_{05} - z_{50}}{z_{50} - z_{95}}, 
\end{equation}
where $z_{05}$ and $z_{95}$ are the redshifts of 5\% and 95\% ionization, respectively.  Note that $A_z =1 $ corresponds to a symmetric history, where the first and second halves of reionization take place over the same redshift interval, and $A_z > 1$ corresponds to a longer first half. The mass-weighted reionization history is given by

\begin{equation}
   \langle x_{\rm HII} \rangle_{\rm M} = \rm{exp} \left[ - \rm{max} \left( \frac{z-a_w}{b_w},0 \right)^{c_w}\right], 
\end{equation}
where we solve for $a_w$, $b_w$, and $c_w$ for given $z_{50}$, $\Delta z_{90}$ and $A_z$ following the prescription in Appendix A of Ref. \cite{Trac2021-xm}.  We adopt wide uniform priors of $z_{50} \in [5,25]$, $\Delta z_{90} \in [0.5, 25]$, and $A_z \in [0.5, 9]$.  As pointed out by Ref. \cite{Millea2018-xl} and emphasized in Ref. \cite{Ilic2025-xj}, taking uniform priors for the parameters describing a reionization history does not in general yield a uniform prior for $\tau_{\rm es}$. This can be an important consideration for interpreting $\tau_{\rm es}$, especially when comparing between different models of the reionization history. Our choice of priors is motivated by the goal of this paper; to understand how the combination of forest and CMB observations constrain the reionization history itself.

\begin{figure}
    \centering    \includegraphics[width=1\linewidth]{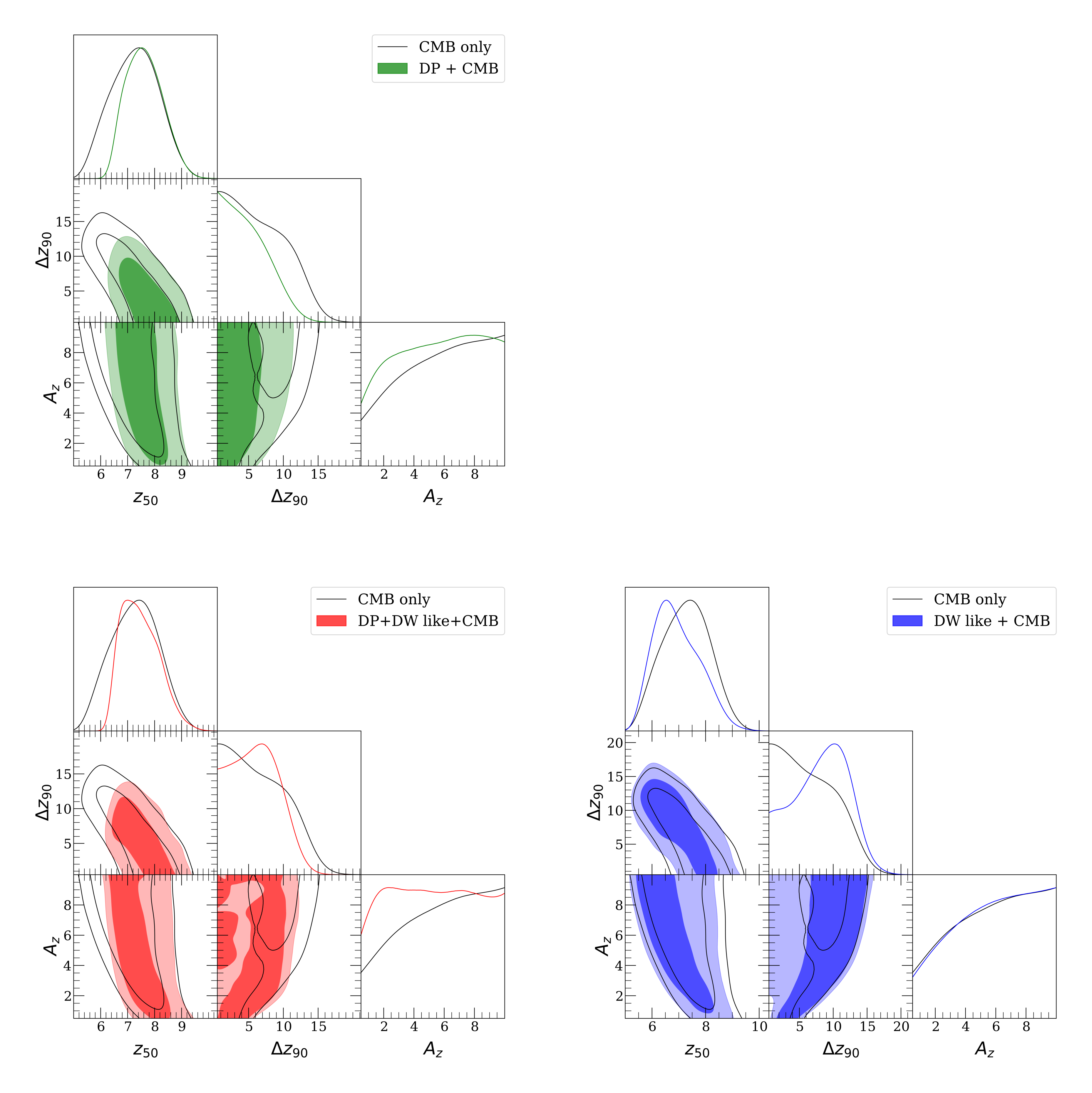}
    \caption{  Constraints on reionization midpoint, duration, and asymmetry from joint CMB-forest analyses using the parameteric reionization model of \cite{Trac2021-xm}. Each triangle-plot corresponds to a different combination of forest observations as indicated in the legends.  In all plots, black curves and contours correspond to the \textsc{CMB Only} baseline. Inner and outter contours correspond to the 68\% and 95\% credible regions, respectively. In the legends, \textsc{DP} refers to upper limits on the volume-weighted mean neutral fraction over $4.9 \lesssim z \lesssim 6.7$ from dark pixel counts in the Lyman-series forests \cite{Davies2025-lf,Jin2023-bh}, while \textsc{DPlike} refers to a lower limit at $z=5.8$ derived from damping-wing-like features on the boundaries of stacked dark gaps in the coeval Ly$\alpha$ and Ly$\beta$ forests \cite{Zhu2024-aq}.}
    \label{fig:Weibull_tri}
\end{figure}

In Figure \ref{fig:Weibull_tri}, we show the posteriors for $z_{50}$, $\Delta z_{90}$ and $A_z$ for different combinations of forest constraints (colored and filled contours), against the baseline \textsc{CMB Only} analysis (black contours).  The top-left and bottom-right sets of panels correspond to the \textsc{DP + CMB} and \textsc{DWlike+CMB} analyses, respectively. The bottom-left set corresponds to the full \textsc{DP+DWlike+CMB} analysis.  The inner and outer contours correspond to the 68\% and 95\% credible regions, respectively.    

We begin with the \textsc{CMB Only} constraints (which are the same in all sets of panels). The marginalized posterior distribution for the reionization midpoint ($z_{50}$; top-left) peaks near $\sim 7.3$, consistent with the constraints from previous works \cite{Planck18}.  The middle-left panel shows an anti-correlation between $z_{50}$ and $\Delta z_{90}$, which arises from a degeneracy in $\tau_{\rm es}$.  Longer durations (higher $\Delta z_{90}$) can be offset by later midpoints (lower $z_{50}$), and conversely, to maintain approximately the same $\tau_{\rm es}$.  We also see that low $z_{50}$ broadly favors a larger asymmetry parameter, $A_z$.  This is related to our requirement that reionization be completed no later than $z=4$.  For models with later midpoints and longer durations, a larger asymmetry is the only way to finish reionization on time.  

Consider now the combination of \textsc{DP+CMB} constraints (top-left set of panels).  The inclusion of dark pixel information disfavors $z_{50}\lesssim6.5$, a result of the upper limit $\langle x_{\rm HI}\rangle_V \leq 39.6\% + 9.3\% (1\sigma)$ at $\bar{z}=6.043$ from Ref \cite{Davies2025-lf}.  The middle panel also shows a tightening of the upper limit on $\Delta z_{90}$.  This results from two effects: (1) As noted above, a longer duration is correlated with a later midpoint owing to their degeneracy in $\tau_{\rm es}$. Therefore, longer durations are disfavored when the $z\sim 6$ dark pixel constraints disfavor the latest midpoints; (2) The dark pixel constraints require $\langle x_{\rm HI} \rangle_V \lesssim 13\%$ at $z\lesssim 5.5$.  All else being equal, this has the effect of requiring the parametric reionization histories to complete earlier than our $z=4$ hard cut-off, resulting in shorter durations.  

The bottom-right panels of Figure \ref{fig:Weibull_tri} show results from the \textsc{DWlike+CMB} combination. Recall that the lower limit on the neutral fraction from Ref. \cite{Zhu2024-aq} complements the upper limits obtained from dark pixel measurements, so it is interesting to consider how this combination differs from the \textsc{DP+CMB} analysis.   We see that the $z_{50}$ posterior is shifted downward and $\Delta z_{90}$ now peaks at $\Delta z_{90}\sim10$.  These extended but delayed histories are able to satisfy the constraint from \cite{Zhu2024-aq} that $\langle x_{\rm HI}\rangle_V(z=5.8)>6.1\%\pm3.8\%$ while also satisfying the CMB limits on $\tau_{\rm es}$.  We note, however, that the effect of the $x_{\rm HI}$ limit is relatively weak. The likelihood saturates at $x_{\rm HI}=0$ at a modest $\sim1.6 \sigma$. Hence, it does not rule out reionization histories that complete earlier. Keeping in mind the fixed-$\tau_{\rm es}$ degeneracy, this accounts for the low-$\Delta z_{90}$ tail seen in the middle panel.  

Finally, the lower-left panels show results from the full combination of constraints, \textsc{DP+DWlike+CMB}. Here we see a steep decline in the posterior for $z_{50} \lesssim 6.5$ from the dark pixel constraints, as well as a tilt towards lower-$z_{50}$ from the damping-wing-like constraint.  The $\Delta z_{90}$ posterior exhibits a peak as in the \textsc{DWlike+CMB} case, but is somewhat weaker.  We note that $A_z$ is poorly constrained (bottom-right).  In the $A_z$--$\Delta z_{90}$ plane (bottom-middle), the posterior now exhibits a multi-peaked structure. Following the discussion above, this structure arises from the competing preferences of the dark-pixel and damping-wing-like constraints. In combination with the CMB, the former favors comparatively rapid completion of reionization, while the latter favors a more extended history with significant late-time neutral fraction. As a result, the \textsc{DP+DWlike+CMB} analysis yields disjoint parameter combinations yielding similar posterior probabilities.  In $\Delta z_{90}-A_z$ space (bottom-middle), the peaks at $\Delta z_{90}\sim 1$ correspond to models in which reionization ends before $z\approx 6.4$, saturating the weak lower limit on the neutral fraction from the damping-wing-like measurement at $z=5.8$ (see Figure \ref{fig:FlexKnot_hist_contour}). This region of the parameter space would be ruled out if future observations are able to strengthen lower limits on the neutral fraction.   

In Table \ref{tab:parameters} we summarize our constraints on $z_{50}$ and $\Delta_{\rm 90}$.  We report the marginalized posterior mean and 68\% credible intervals on the former and 95\% upper limits on the latter. In the full joint CMB-forest analysis, we find 

\begin{equation}
z_{50}=7.5^{+0.5}_{-0.9}
\end{equation}
\begin{equation}
\Delta z_{90}<11.5~~(95\%~\mathrm{upper~limit}).
\end{equation}
The latter is complementary to recent constraints on $\Delta z_{90}$ from the patchy kinetic Sunyaev Zel'dovich (pkSZ) effect \cite{Beringue2025-ka,Chaubal2026-wu, Reichardt2020}. Our upper limit is weaker than the 95\% limits of $\Delta z_{90}<7.7$ obtained from ACT DR6 \cite{Beringue2025-ka} and $\Delta z_{90}<7.0$ from the recent SPT-3G analysis \cite{Chaubal2026-wu}. These rely on scaling relations calibrated using semi-numerical AMBER simulations \cite{Kramer2025}. Additionally, the reionization midpoint is fixed to $z_{50}=8$ in the ACT analysis and $z_{50}=7.68$ in the SPT-3G analysis, while the adopted AMBER scaling relation is calibrated at fixed asymmetry $A_z=3$, rather than jointly marginalizing over $z_{50}$, $\Delta z_{90}$, and $A_z$.

We also tabulate our derived constraints on $\tau_{\rm es}$ and 95\% upper limits on the $z>15$ contribution to the optical depth, $\tau_{\rm es}^{z>15}$.  We remind the reader of the implicit non-uniform prior for $\tau_{\rm es}$ that arises from adopting uniform priors on the Weibull function parameters. The inferred total optical depth is stable across the different likelihood combinations, with all central values agreeing to much better than 1$\sigma$ and with essentially no change in the uncertainty. The forest constraints therefore primarily redistribute the allowed optical depth in redshift rather than altering its total value. The dark pixel data favor a slightly larger contribution from lower redshift while suppressing an extended high-redshift tail, whereas the damping-wing-like data permits a larger $z>15$ contribution while shifting the total optical depth slightly downward. When both forest likelihoods are included, their opposing preferences largely cancel in $\tau_{\rm es}$, although the upper limit on $\tau_{\rm es}^{z>15}$ remains tighter than for the CMB alone.

\subsection{\textsc{FlexKnot Constraints}}
\label{sec:FlexKnot}

We now consider a more flexible model for the reionization history. In this section, we use the \textsc{FlexKnot} approach first applied in this context by Ref. \cite{Millea2018-xl}.  In this approach, the reionization history is defined by a number of $(z_i, x_i^{\rm HII})$ pairs called knots. The reionization history between the knots is interpolated with piecewise cubic Hermite interpolating polynomials (PCHIP).  We enforce monotonicity in the knots.  In this case, the use of PCHIP ensures that the resultant reionization history will also be monotonic.  Our fiducial \textsc{FlexKnot} model has four free parameters: $z_{\rm beg}$, $z_{\rm end}$, $z_1$ and $x_1$, where $z_{\rm beg}$ and $z_{\rm end}$ denote the redshifts at which $x_{\rm HII} = 0$ and $1$, respectively, and the $(z_1 , x_1)$ pair determines the position of the free-floating knot between $z_{\rm beg}$ and $z_{\rm end}$.\footnote{Note that this is different than the original \textsc{FlexKnot}s used by Ref. \cite{Millea2018-xl}, which instead had fixed nodes at the start and end of the allowed redshift range.}  This model is similar to the \textsc{3-flex} model used in Ref. \cite{Ilic2025-xj}.  We adopt wide priors of { $z_{\rm end}$, $z_1$, $z_{\rm beg}$ } $\in [4,30]$, and $0 \le x_1 \le1$ but construct our likelihoods to ensure $z_{\rm end} < z_1 < z_{\rm beg}$. In Appendix \ref{sec:PR3_vs_PR4}, we present our constraints on the cosmological parameters, which we find are stable across the different dataset combinations and reionization modeling methods applied in this paper.  

\begin{figure}
    \centering
    \includegraphics[width=0.98\linewidth]{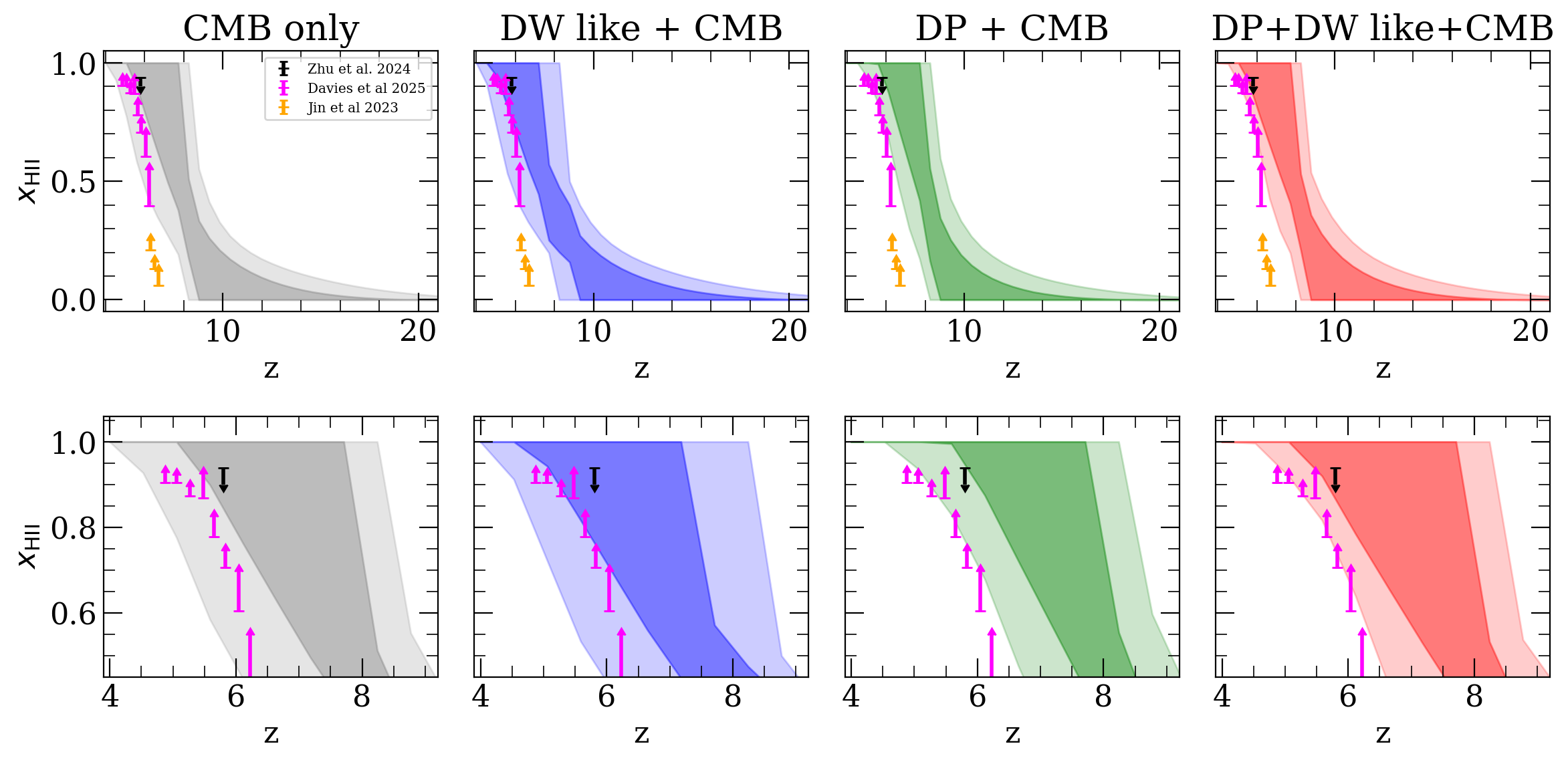}
    \caption{Constraints on the global neutral fraction as a function of redshift from joint CMB-forest analyses using \textsc{FlexKnots}. Each column corresponds to a different combination of forest observations, as annotated above. The bottom row shows a zoomed-in version of the top row, highlighting the second half of reionization and the effect of the forest constraints. For reference, the magenta, yellow and black arrows mark the $1\sigma$ constraints from Refs. \cite{Davies2025-lf}, \cite{Jin2023-bh} and \cite{Zhu2024-aq}, respectively.   }
    \label{fig:FlexKnot_hist_contour}
\end{figure}

Figure \ref{fig:FlexKnot_hist_contour} shows the marginalized constraints on the reionization history for the different dataset combinations. From left to right, the columns correspond to \textsc{CMB Only}, \textsc{DWlike+CMB}, \textsc{DP+CMB} and \textsc{DP+DWlike+CMB}, respectively.  The top row shows the entire reionization history, while the bottom row zooms in on the second half of reionization to highlight the impact of the forest constraints in that epoch.  For reference, each panel also displays all of the dark pixel and damping-wing-like constraints used in our inference pipeline (although each column uses a different set of these).  The inclusion of the damping-wing-like lower limit on $\langle x_{\rm HI}\rangle_V$ has a modest effect on the reionization history contours, tightening the 1$\sigma$ limits only slightly relative to the \textsc{CMB Only} case.  On the other hand, the addition of the dark pixel constraints has a strong impact on the contours at $z\lesssim6.5$, requiring reionization to be more than $\sim 80\%$ complete by $z \sim 6$.

\begin{figure}
    \centering
    \includegraphics[width=1\linewidth]{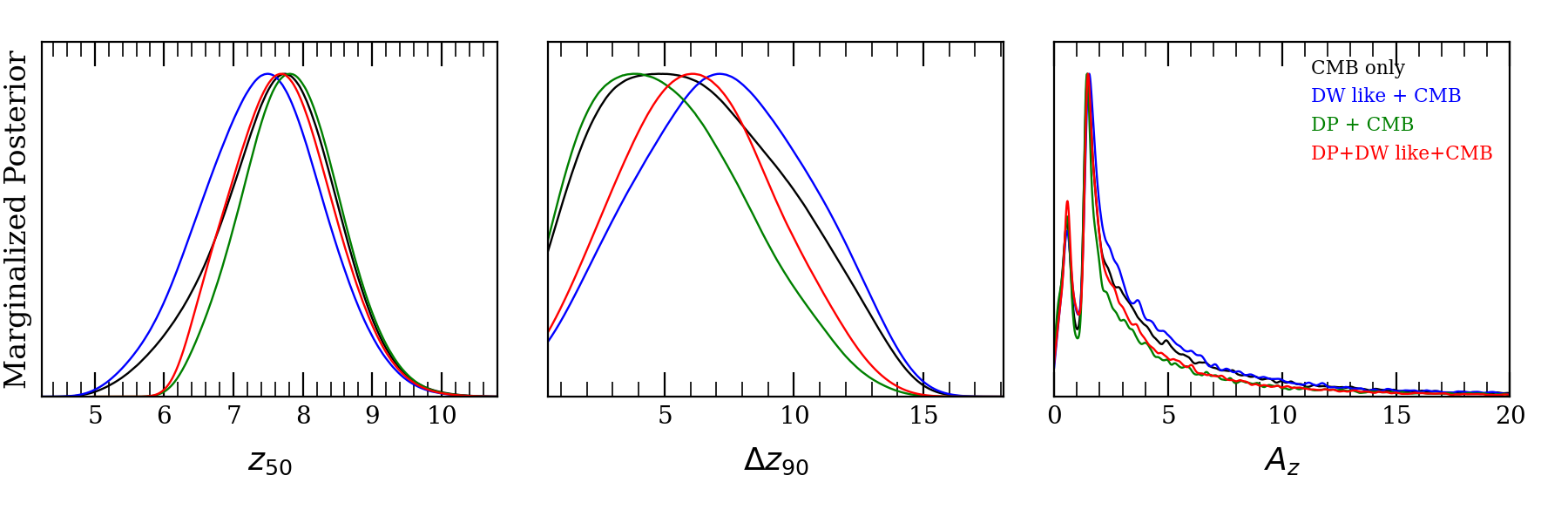}
    \caption{Derived constraints on the midpoint, duration, and asymmetry of reionization from joint CMB-forest analyses using \textsc{FlexKnots}.  Different colors represent different combinations of forest data, as indicated in the legend. In the \textsc{DWlike+CMB} analysis, the lower limit on the neutral fraction at $z=5.8$ results in a boosted preference for more extended reionization histories with lower-$z_{50}$ relative to the baseline \textsc{CMB Only} analysis.  In the \textsc{DP+CMB} analysis, the upper limits from the dark pixel measurements exclude low-$z_{50}$, producing a preference for shorter durations in order to preserve approximately the same $\tau_{\rm es}$. In a full joint analysis (\textsc{DP+DWlike+CMB}), these competing effects yield an improved constraint on reionization's duration.}
    \label{fig:Deltaz_FlexKnot_corner}
\end{figure}

\begin{figure}
    \centering
    \includegraphics[width=0.95\linewidth]{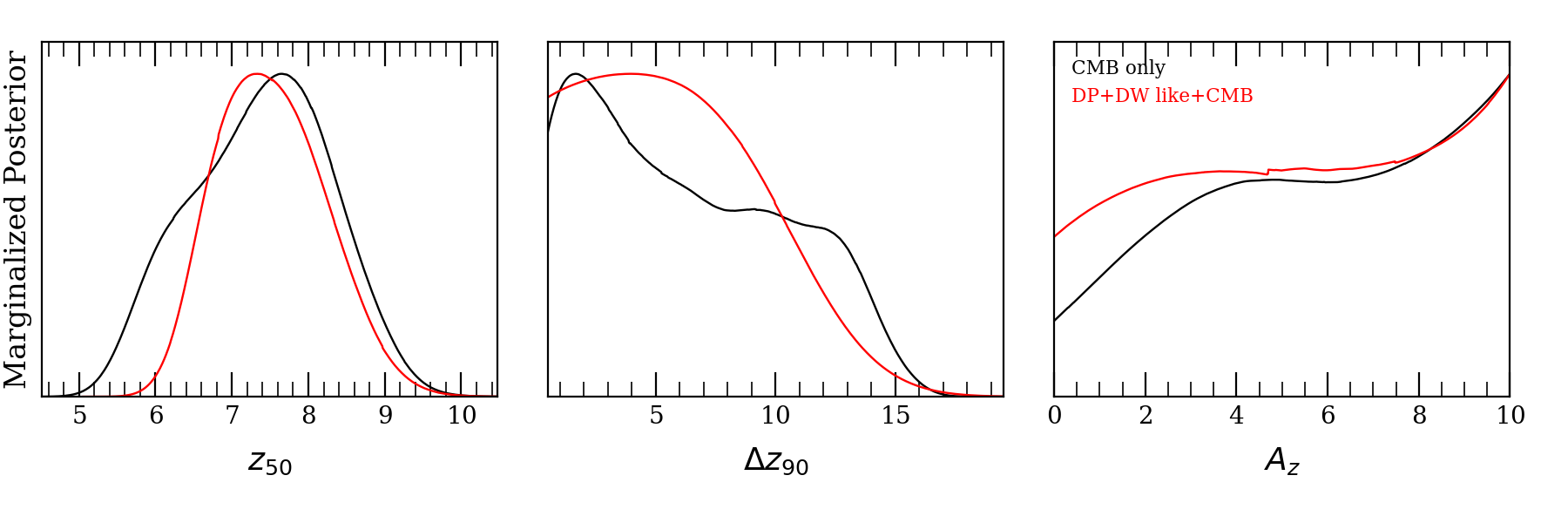}
    \caption{Marginalized posteriors on $z_{50}$, $\Delta z_{90}$, and $A_z$ from our \textsc{FlexKnots} analyses after jointly flattening the implicit priors on these parameters using importance sampling. These results should be compared to Figures \ref{fig:Weibull_tri} and \ref{fig:Deltaz_FlexKnot_corner}. After flattening the priors, there is little change in the $z_{50}$ posteriors.  The $\Delta z_{50}$ and $A_z$ posteriors, however, are more sensitive to the adopted priors.}
    \label{fig:priors}
\end{figure}

In Figure \ref{fig:Deltaz_FlexKnot_corner}, we plot posteriors for the derived parameters $z_{\rm 50}$, $\Delta z_{90}$, and $A_z$. The middle section of Table \ref{tab:parameters} summarizes the constraints.  Consistent with the results of the last section, the addition of the dark pixel constraints in \textsc{DP+CMB} disfavors $z_{50}\lesssim6.5$. This again leads to a preference for shorter $\Delta z_{90}$ relative to the \textsc{CMB Only} result in order to preserve a roughly constant $\tau_{\rm es}$.  In the \textsc{DWlike+CMB} case, we also see a shift towards lower $z_{\rm 50}$ and longer $\Delta z_{90}$ relative to the \textsc{CMB Only} case.  Comparing the posterior distributions of $z_{\rm 50}$, the main effects of the dark-pixel and damping-wing-like constraints are to pull the distributions toward earlier and latter midpoints, respectively.  These constraints also pull the posterior of $\Delta z_{90}$ toward shorter and longer durations, respectively.  When combined, the net result is a mild reduction in the widths of these distributions. In the full joint CMB-forest analysis, we find 

\begin{equation}
z_{50}=7.7^{+0.7}_{-0.8}
\end{equation}
\begin{equation}
\Delta z_{90}<11.4~~(95\%~\mathrm{upper~limit}).
\end{equation} 

The posterior for $A_z$ exhibits two interesting features. First, the spike at $A_z<1$ owes to the definition of $A_z$.  When sampling over the uniform priors of the \textsc{FlexKnot} parameters, the samples are evenly split between $A_z>1$ and $A_z<1$.  This means that half of the samples are constrained to $0<A_z<1$ and the other half to $A_z > 1$.  Thus, the sample density with $A_z<1$ is much higher, resulting in the observed spike.  Secondly, the $A_z$ posterior is very different from the one obtained in the parametric analysis (compare to the bottom-right panels of each triangle plot if Fig. \ref{fig:Weibull_tri}).  This owes to $A_z$ not being well constrained by the data and the different effective priors between the analyses.  In the parametric analysis, we adopted flat priors on $z_{50}$, $\Delta z_{90}$, and $A_z$, while in the \textsc{FlexKnot} analysis we used flat priors for $z_{\rm beg}$, $z_{\rm end}$, $z_1$, and $x_1$.  

\begin{table}
    \centering
    \begin{tabular}{l|c|c|c|c|c}
        \hline
        Parametric (Weibull) & $z_{50}$ & $\Delta z_{90}$ & $A_z$ & $\tau_{\rm es}$  & $\tau_{\rm es}^{z>15}$   \\
         &  &   &  &   \\
        [0.2ex]
        \hline
        \textsc{CMB Only}       & $7.3\pm0.9$         & $<13.5$ & - & $0.0607 \pm 0.0063$ & < 0.0075 \\ [0.5ex]
        \textsc{DWlike+CMB}     & $6.9^{+0.7}_{-1.1}$ & $<14.2$ & - & $0.0602 \pm 0.0061$ & < 0.0084 \\ [0.5ex]
        \textsc{DP+CMB}         & $7.6^{+0.6}_{-0.8}$ & $<10.6$ & - & $0.0612^{+0.0060}_{-0.0065}$ & < 0.0048 \\ [0.5ex] 
        \textsc{DP+DWlike+CMB}  & $7.5^{+0.5}_{-0.9}$ & $<11.5$ & - & $0.0611 \pm 0.0062$ & < 0.0058 \\ [0.5ex]
      \hline
\multicolumn{6}{l}{\textsc{FlexKnot}} \\[0.2ex]
\hline
\textsc{CMB Only}\rule{0pt}{3.2ex}
    & $7.6^{+0.9}_{-0.7}$
    & $<12.3$
    & -
    & $0.0608^{+0.0058}_{-0.0065}$
    & $<0.0044$ \\[0.5ex]
         \textsc{DWlike+CMB}    & $7.4\pm0.9$           &  $<12.8$ & - & $0.0600\pm0.0062$            & $< 0.0048$ \\ [0.5ex]
         \textsc{DP+CMB}        & $7.8\pm0.7$           &  $<10.6$ & - & $0.0612^{+0.0058}_{-0.0064}$ & $< 0.0032$ \\ [0.5ex]
         \textsc{DP+DWlike+CMB} & $7.7^{+0.7}_{-0.8}$   &  $<11.4$ & - & $0.0610^{+0.0058}_{-0.0065}$ & $< 0.0039$ \\ [0.2ex]
    \hline
\multicolumn{6}{l}{\textsc{Semi-analytic} (fixed cosmology and IGM opacity model)} \\ [0.2ex]
\hline
\textsc{CMB Only}\rule{0pt}{3.2ex}  & $7.4 \pm 0.3$         & $<7.8$   & - & $0.0564\pm0.0024$ & $<0.0011$ \\ [0.5ex]
\textsc{DP+DWlike+CMB}              & $7.4 \pm 0.3$         & $<7.6$   & - & $0.0562\pm0.0024$ & $<0.0011$ \\ [0.5ex]
\textsc{$\Gamma_{\rm HI}$ + CMB}    & $7.4^{+0.2}_{-0.3}$   & $6.0^{+0.8}_{-1.0}$   & $2.4^{+0.3}_{-0.7}$ & $0.0564\pm0.0024$ & $<0.0011$ \\ [0.2ex]
    \end{tabular}
    \caption{Constraints on reionization history summary statistics from joint CMB-forest analyses using a parameteric reionization model (top), \textsc{FlexKnots} (middle), and the semi-analytic model of Ref. \cite{Cain2026-op} (bottom). We report marginalized posterior means with 68\% credible intervals, and 95\% upper limits where appropriate. In the right-most column, we report 95\% upper limits on the $z>15$ contribution to the optical depth.  For the semi-analytic model results (bottom 3 rows), the modest downward shift in $\tau_{\rm es}$ and the reduction in its uncertainties owe to the cosmological parameters being held fixed in that analysis (see \S \ref{sec:results_Gamma}). 
    }
    \label{tab:parameters}
\end{table} 

In Figure \ref{fig:priors}, we explore the effects of implicit priors on our marginalized posteriors for $z_{50}$, $\Delta z_{90}$, and $A_z$ using importance sampling. We obtained the implicit prior distribution $\mathcal{P}(z_{50}, \Delta z_{90}, A_z)$, by sampling over the uniform prior distributions of $z_{\rm beg}$, $z_{\rm end}$, $z_1$, and $x_1$.  We then assigned weights of $\mathcal{P}(z_{50}, \Delta z_{90}, A_z)^{-1}$ to each sample (see e.g. \cite{Ilic2025-xj}).   Figure \ref{fig:priors} shows the resulting fully marginalized posteriors for the \textsc{CMB Only} (black) and \textsc{DP+DWlike+CMB} analyses.  While there is little change in the $z_{50}$ posteriors, the $\Delta z_{50}$ and $A_z$ posteriors are qualitatively more similar to those obtained in the parametric model analyses (Fig. \ref{fig:Weibull_tri}). Notably, the peak in the $\Delta z_{90}$ posterior within the \textsc{DP+DWlike+CMB} analysis is significantly weaker, illustrating that the implicit prior on $\Delta z_{90}$ has an important effect on the shape of this distribution.  The large difference in the $A_z$ posterior also confirms our assessment in the previous paragraph.  After flattening the priors, we find $z_{50} = 7.4^{+1.1}_{-0.9}$, $\Delta z_{90}<13.6$ for \textsc{CMB Only}, and $z_{50}=7.5^{+0.6}_{-0.8}$, $\Delta z_{90}<12.1$ for \textsc{DP+DWlike+CMB}.      

The top two sections of Table \ref{tab:parameters} show that the constraints on $\tau_{\rm es}$ are remarkably stable across not only the different combinations of data sets, but also across the parametric and \textsc{FlexKnot} modeling methods.  The 95\% limits on the $z>15$ contribution to the optical depth, however, do depend on the choice of modeling method, with our implementation of $\textsc{FlexKnot}$ yielding tighter upper limits than the parametric model.  Again, at least part of these differences may owe to the different priors adopted between the two modeling methods.  At a qualitative level, both analyses show that combining the CMB with information on the end of reionization tightens limits on the earliest stages of reionization.

\section{Tightening Constraints with the Ly$\alpha$ Forest Transmission}
\label{sec:results_Gamma}

In the preceding sections, we used dark-pixel and damping-wing-like absorption measurements to bracket the neutral fraction during the final stages of reionization. These relatively direct probes, however, do not exhaust the information contained in the Lyman-series forests. In particular, the redshift evolution and sightline-to-sightline fluctuations of the Ly$\alpha$ forest transmission are highly sensitive to the timing of reionization's end. For example, forward-modeling the distribution of forest effective optical depths, $\tau_{\rm eff}$, can yield substantially tighter constraints on the reionization history \citep[e.g.,][]{2021MNRAS.506.2390Q,2025PASA...42...49Q}. Such an analysis requires a detailed model of the density, temperature, and ionizing background fluctuations that shape the forest, and lies beyond the scope of this paper. To illustrate the power of exploiting the additional information embedded within the forest, we instead use measurements of the hydrogen photoionization rate, $\Gamma_{\rm HI}$, which essentially encodes the evolution of the mean Ly$\alpha$ transmission. Although this approach does not capture the full constraining power of forest statistics, it demonstrates how strongly the mean transmission alone can sharpen the constraints. It also motivates the development of accurate and computationally efficient models for future analyses.

\subsection{Methods}

To connect the mean-transmission measurements to the reionization history, we use the semi-analytic model of Ref. \cite{Cain2026-op} which is derived from the cosmological radiative transfer equation.    Given a source ionizing emissivity history, the model jointly solves for the mass-weighted ionized fraction, $x_{\rm HII}(z)$, and $\Gamma_{\rm HI}(z)$, accounting for absorptions both within ionized regions and at the boundaries of the neutral IGM. In the model, these absorptions are accounted for with the total absorption coefficient such that the mean free path of ionizing photons can be written $\lambda^{-1} = \lambda_n^{-1}+\lambda_i^{-1}$.  Here, $\lambda_n$ is the average distance ionizing photons travel to reach the edges of bubbles\footnote{This distance is often taken to be a definition of the bubble size.}, and $\lambda_i$ is the mean free path through ionized gas inside the bubbles. For the former, we measure the mean distance to neutral regions for randomly positioned skewers through the ``Late start/Late end'' reionization simulation of Ref. \cite{Cain2024-wt}. We find that this mean distance is well-fit by the following form 

\begin{equation}
    \label{eqn:lambda_N}
    \lambda_n(x_{\rm HII}) = A \frac{e^{B x_{\rm HII}}} {(1-x_{\rm HII})},
\end{equation}
where $A=0.278$ physical Mpc and $B=2.060$. Absorptions within the reionized IGM are quantified by an effective clumping factor, termed the absorption rate clumping factor,

\begin{equation}
    \label{eqn:clumping}
    C_{\rm A} = \frac{\dot{n}_{\rm abs}^{\rm ionized}}{\alpha(T_{\rm ref}) (1+\chi)n_{\rm H}^2 x_{\rm HII}},
\end{equation}
where $\dot{n}_{\rm abs}^{\rm ionized}$ is the ionizing photon absorption rate density in the gas, $\alpha_{\rm B}(T_{\rm ref})$ is the case-B recombination coefficient evaluated at a reference temperature, $T_{\rm ref} = 10,000$ K, and the factor $1+\chi = 1 + n_{\rm H}/n_{\rm He} \approx 1.08$ (where $n_H$ and $n_{\rm He}$ are the hydrogen and helium number densities, respectively) accounts for electrons from singly ionized helium.\footnote{We adopt the widely used assumption that helium is singly ionized at the same time as hydrogen. The second reionization of helium likely occurred later in the cosmological timeline, driven by the rise of quasar population \cite[e.g.][]{2016MNRAS.460.1885U, Gaikwad2025}.}  Motivated by results from high-resolution radiative hydrodynamics simulations \cite{2020ApJ...898..149D, 2021ApJ...923..161N, 2026arXiv260325788C}, we adopt the power-law ansatz $C_{A}=3 \times \left(\Gamma_{\rm HI}/3\times10^{-13}\text{s}^{-1} / x_{\rm HII} \right)^{1-\xi}$.  Since $\lambda_i \propto \Gamma_{\rm HI}/C_A$, this scaling is equivalent to $\lambda_i\propto\Gamma_{\rm HI}^{\xi}$.  Here we use the widely adopted value of $\xi = 2/3$ \cite[see e.g.][]{2011ApJ...743...82M, 2016MNRAS.460.1328D, DAloisio2018, 2023MNRAS.525.4093G}. In principle, the parameters describing $\lambda_n$ and $\lambda_i$ could be incorporated into the inference and marginalized over. To reduce computational cost for this illustrative analysis, we leave the parameters fixed to their simulation-motivated values.  We parameterize the comoving ionizing emissivity with a double power law,

\begin{equation}
    \label{eqn:DPL}
    \dot{n}_{\rm ion}(z) = \frac{N_*}{ \left(\frac{1+z}{1+z_*}\right)^{-\alpha_*} + \left(\frac{1+z}{1+z_*}\right)^{-\beta_*} },
\end{equation}
where, $N_*$, $z_*$, $\alpha_*$ and $\beta_*$ are the free parameters that we sample over. We take uniform priors of log($N*$) $\in [-23, -20]$, $z_*\in [5, 12]$, $\alpha_*\in[-8,0]$, $\beta_* \in [-6,6]$, and require (without loss of generality) that $\alpha_*<\beta_*$. In this case, $\alpha_*$ sets how quickly the ionizing emissivity grows at $z>z_*$, and $\beta_*$ controls the evolution at $z < z_*$.

We use measurements of the spatially averaged hydrogen photoionization rate, $\langle\Gamma_{\rm HI}\rangle$, from Ref. \cite{Gaikwad2023-mm}. These measurements are inferred from the Ly$\alpha$ forest transmission in a sample of 67 quasar spectra spanning $4.9\lesssim z\lesssim6.0$. Ref. \cite{Gaikwad2023-mm} forward-modeled the observations by post-processing hydrodynamic simulations with spatially fluctuating ionizing radiation fields and extracting mock Ly$\alpha$ forest spectra. They then compared the distribution of effective optical depths, $\tau_{\rm eff}$, between the observed and mock spectra to jointly constrain $\langle\Gamma_{\rm HI}\rangle$ and the ionizing-photon mean free path. The reported uncertainties account for the thermal state of the IGM, modeling assumptions, cosmic variance, and observational uncertainties such as noise, sky subtraction, and continuum modeling.  In this work, we use only the marginalized measurements of $\langle\Gamma_{\rm HI}\rangle$. Our analysis therefore uses essentially the redshift evolution of the mean Ly$\alpha$ forest transmission, but does not attempt to model and leverage sightline-to-sightline fluctuations in the transmission. 

For simplicity, we approximate the likelihood for the $\Gamma_{\rm HI}$ measurements with piecewise-asymmetric Gaussians and assume that the redshift bins are independent.  For each of the 12 redshift bins, the width on either side of the mean is the relevant $1\sigma$ upper or lower measurement error. This approach neglects correlations between redshift bins arising from shared modeling assumptions and observational systematics in the inference of $\Gamma_{\rm HI}$.  To assess the effects of neglecting such correlations, we performed two tests in which we retained only four and one of the $\Gamma_{\rm HI}$ measurements in our analysis (see Appendix \ref{sec:Gam_like_test}).  We find that the posteriors in the four-redshift test are very similar to those in our fiducial analysis (with 12 redshift bins). Notably, the fully marginalized posterior for $\Delta z_{90}$ remains similar even in the one-redshift test run.  Based on these findings, we argue that our results and conclusions would likely be similar in a full analysis that accounts for the covariance between $\Gamma_{\rm HI}$ redshift bins (all other details remaining the same).   Because of the computational cost of solving differential equations in our semi-analytic model millions of times, and the approximate nature of our $\Gamma_{\rm HI}$ likelihood, in this section we keep the cosmological parameters fixed at: $A_s=2.1\times10^{-9}$, $n_s=0.965$, $h=0.674$, $\Omega_b h^2=0.022$, $\Omega_c h^2=0.12$, consistent with Planck measurements \cite{Planck18, Tristram2024}.\footnote{This is equivalent to ln($10^{10} A_s)\approx3.045$}

\subsection{Results}

\begin{figure}
    \centering
    \includegraphics[width=0.95\linewidth]{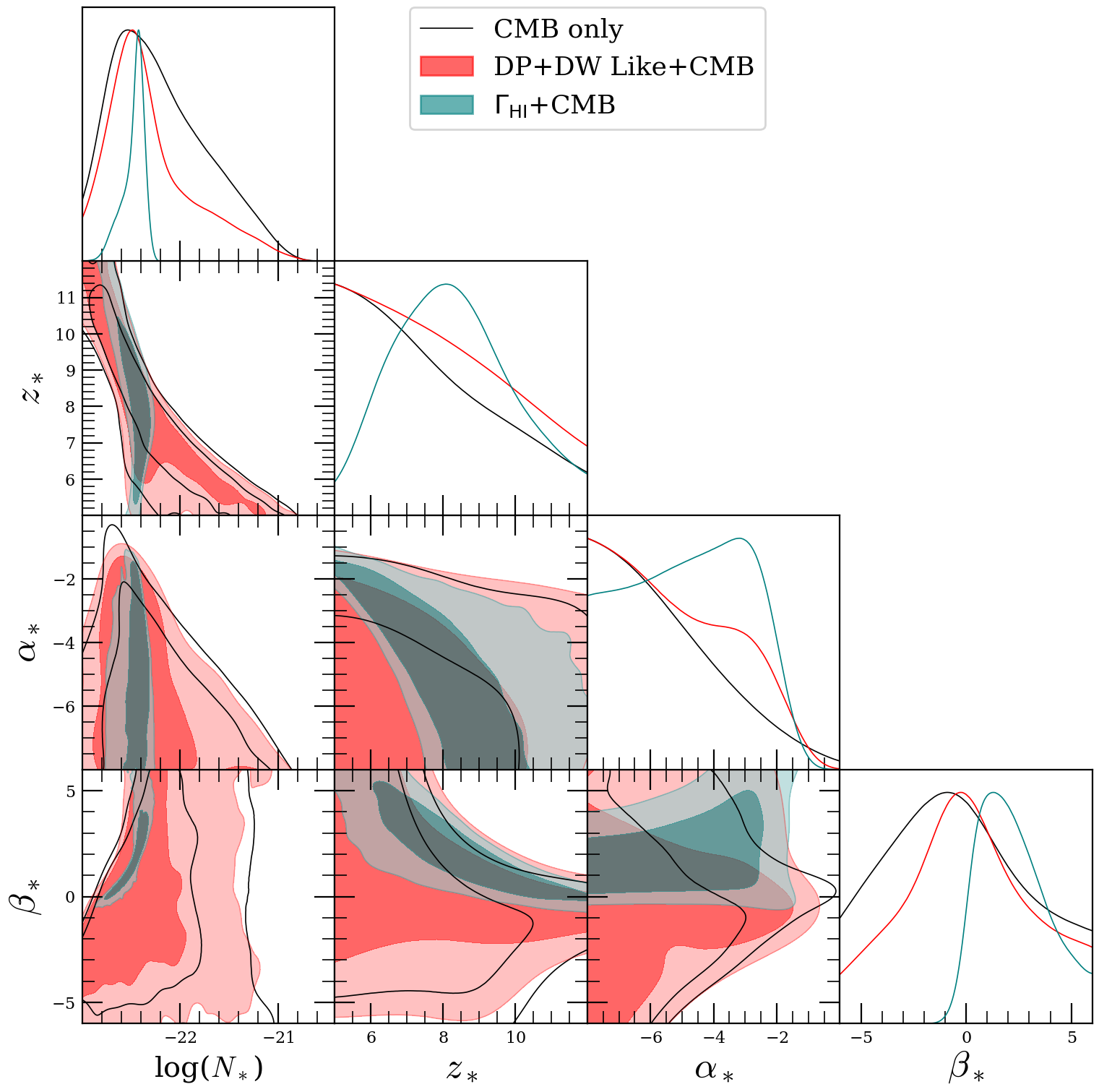}
    \caption{ Constraints on the ionizing emissivity of reionization sources from joint CMB-forest analyses. The emissivity is modeled with a double-power-law with 4 free parameters.  The high and low redshift logarithmic slopes of the emissivity are set by $\alpha_*$ and $\beta_*$, respectively.  Note that positive values of $\beta_*$ correspond to an emissivity that declines toward lower redshift.  The turnover redshift and overall normalization of the emissivity are set by $z_*$ and log($N_*$), respectively.  The \textsc{$\Gamma_{\rm HI}$+CMB} analysis applies measurements of the hydrogen photoionization rate of the IGM at $4.9 \lesssim z \lesssim 6$ from Ref. \cite{Gaikwad2023-mm}, which tightly constrain the last half of reionization compared to the dark pixel and damping-wing-like constraints on the neutral fraction.  }
    \label{fig:DPLem_params}
\end{figure}

Figure~\ref{fig:DPLem_params} shows the posterior constraints on the four parameters of the double-power-law emissivity model in 
Eq.~\eqref{eqn:DPL}. The values of $\alpha_*$ and $\beta_*$ determine the asymptotic slopes at $z>z_*$ and $z<z_*$, respectively, while $z_*$ sets the redshift at which the emissivity changes slope and $N_*$ determines its normalization.\footnote{With the convention adopted in Equation~\eqref{eqn:DPL}, $\dot{n}_{\rm ion}(z_*)=N_*/2$.}  For reference, more negative values of $\alpha_*$ correspond to a more rapidly increasing emissivity (with cosmic time) before $z=z_*$, whereas positive values of $\beta_*$ imply that the emissivity declines toward lower redshift below the break.

The \textsc{CMB Only} posterior exhibits substantial degeneracies among the emissivity parameters. In particular, $N_*$ and $z_*$ are strongly anticorrelated: models in which the break occurs at lower redshift require a larger normalization to produce sufficient ionization at earlier times. These larger normalizations are in turn associated with more negative 
values of $\alpha_*$, which suppress the emissivity at $z>z_*$ and prevent excessive early ionization. Adding the dark-pixel and damping-wing-like constraints removes part of the allowed parameter space, but leaves this degeneracy broadly intact.

The measurements of $\Gamma_{\rm HI}$ substantially sharpen the constraints.  Most notably, they tightly restrict the normalization of the emissivity and remove the high-$N_*$, low-$z_*$ tail that is allowed by the CMB plus neutral-fraction constraints. Models with high-$N_*$ and low-$z_*$ can stay consistent with $\tau_{\rm es}$ by combining a late break with a large normalization and a steep high-redshift slope, but they yield an ionizing background intensity near $z\sim5-6$ that is too large. The \textsc{$\Gamma_{\rm HI}$+CMB} posterior instead favors a break near $z_*\sim8$, together with an emissivity that declines toward lower redshift.  However, significant correlations between $z_*$, $\alpha_*$, and $\beta_*$ remain.

The relative roles of $\alpha_*$ and $\beta_*$ depend on the location of the double power-law break. When $z_* \sim 6$, the $\Gamma_{\rm HI}$ measurements cover a relatively short redshift interval below $z_*$ and therefore provide relatively little leverage on $\beta_*$. In this region of the parameter space, $\alpha_*$ mostly controls the preceding reionization history. For scenarios in which $z_*\sim8$--$9$, the forest measurements lie well below the break and therefore better constrain $\beta_*$. In this case, the value of $\alpha_*$ can vary over a broader range of negative values. These effects can be seen in the degeneracies between $\alpha_*$, $\beta_*$ and $z_*$ in Figure \ref{fig:DPLem_params}. 

The CMB plays a key role in limiting the amount of early ionization. At fixed $N_*$ and $z_*$, making $\alpha_*$ less negative increases the emissivity at high redshift and can produce an excessively large $\tau_{\rm es}$ and low-$\ell$ $E$-mode polarization signal. On the other hand, the forest data anchor the ionizing background near the end of reionization and particularly constrain the evolution of the emissivity near this epoch. The two data sets therefore provide complementary constraints on different 
portions of the emissivity history.

\begin{figure}
    \centering
    \includegraphics[width=0.9\linewidth]{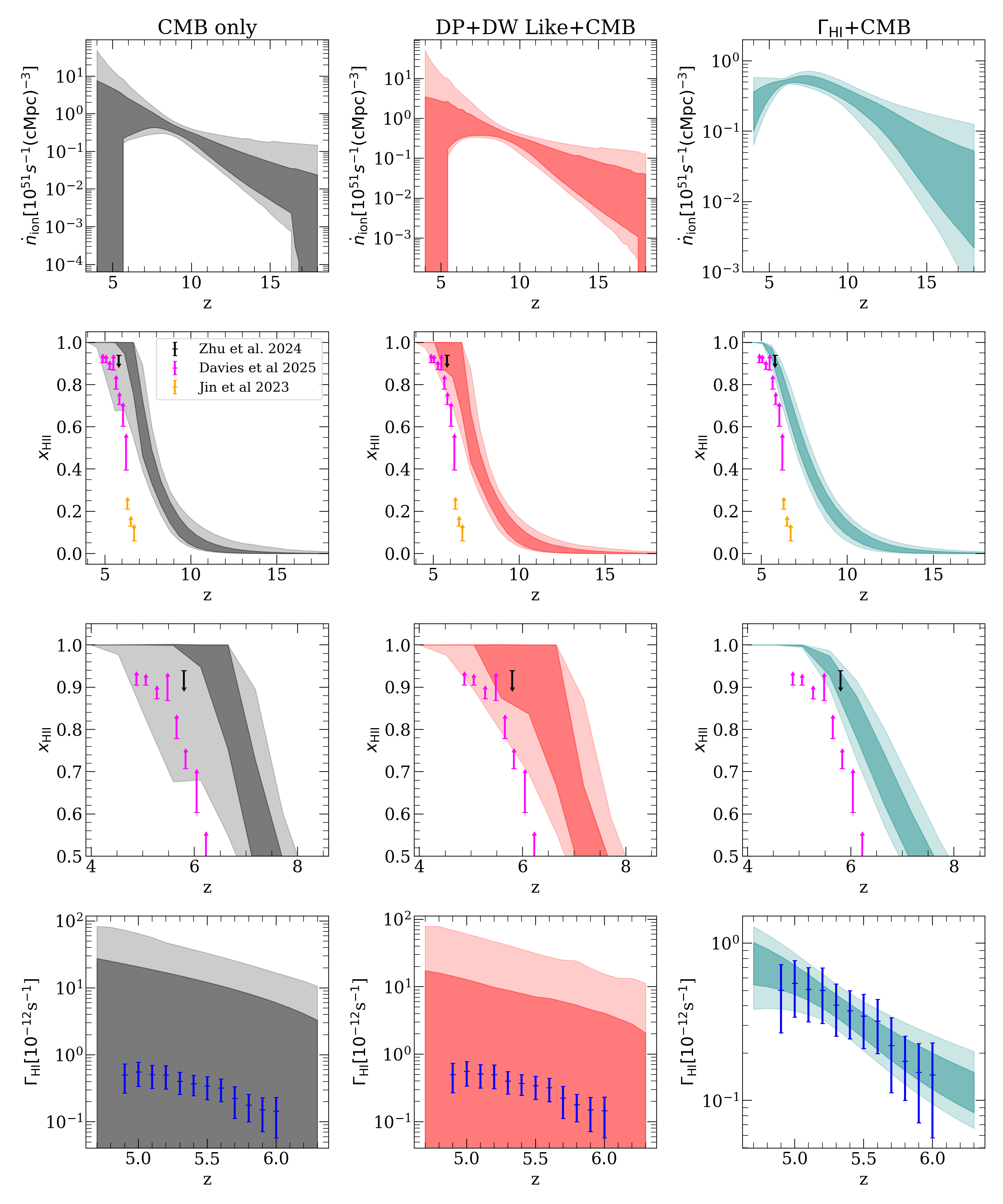}
    \caption{ Constraints on the ionizing emissivity (top), reionization history (middle two), and IGM photoionization rate (bottom) from joint CMB-forest analyses. The 3rd row from the top is a zoomed-in version of the 2nd row that highlights the last half of reionization.   Starting on the left, the columns correspond to the \textsc{CMB Only}, \textsc{DP+DWlike+CMB}, and \textsc{$\Gamma_{\rm HI}$+CMB} analyses, respectively.  For reference, the data points in the bottom row show the $\Gamma_{\rm HI}$ measurements of \cite{Gaikwad2023-mm}. The Ly$\alpha$ forest transmission (from which the $\Gamma_{\rm HI}$ measurements are derived) is highly sensitive to the timing of reionization's end, and provides significantly tighter constraints on reionization compared to the dark pixel and damping-wing-like limits on the neutral fraction.   
    }
    \label{fig:DPL_xHII_hist}
\end{figure}

Figure~\ref{fig:DPL_xHII_hist} illustrates how the constraints on the double-power-law emissivity parameters propagate into the ionizing emissivity, the global reionization history, and the hydrogen photoionization rate. The three columns correspond to the
\textsc{CMB Only}, \textsc{DP+DWlike+CMB}, and \textsc{$\Gamma_{\rm HI}$+CMB} analyses, respectively. From top to bottom, the rows show the comoving ionizing emissivity ($\dot{n}_{\rm ion}$), the ionized fraction ($x_{\rm HII}$), a zoomed-in view of $x_{\rm HII}$ during the final stages of reionization, and the spatially averaged photoionization rate ($\Gamma_{\rm HI}$). The dark and light shaded regions denote the marginalized $68\%$ and $95\%$ credible intervals. In the panels showing $x_{\rm HII}$, for reference we overplot the observational constraints from Refs.~\cite{Zhu2024-aq,Davies2025-lf,Jin2023-bh}, while the lower panels show the $\Gamma_{\rm HI}$ measurements from Ref.~\cite{Gaikwad2023-mm}.

The top row highlights the limited ability of the CMB and the direct neutral-fraction constraints to determine the ionizing
emissivity near the end of reionization. Both the \textsc{CMB Only} and \textsc{DP+DWlike+CMB} posteriors allow a wide range of
emissivities at $z\lesssim 6$. The $\Gamma_{\rm HI}$ measurements substantially reduce this freedom. The \textsc{$\Gamma_{\rm HI}$+CMB} posterior favors an emissivity that rises toward a broad maximum at $z\sim7-8$ and subsequently declines.  This is qualitatively similar to the behavior of the emissivities in reionization simulations that are calibrated to reproduce the evolution of the mean Ly$\alpha$ forest transmission \cite[e.g.][]{2019MNRAS.485L..24K, 2020MNRAS.491.1736K, 2024MNRAS.531.1951C}. There has been some debate as to whether the rapid decline required in the simulations is an artifact of modeling deficiencies \cite[see][]{2024MNRAS.531.1951C}.  Our results provide additional evidence that the emissivity of the reionization source population may in fact decline after $z\sim7-8$. If confirmed, such a decline may owe to a decline in the escape fraction \cite{2022MNRAS.515.2386R, 2022A&A...663A..66M, 2023MNRAS.518..270K} and/or the onset of a feedback process, e.g. suppression of star formation by reionization itself \cite{1992MNRAS.256P..43E, 1994ApJ...427...25S, 2007MNRAS.376..534I, 2021MNRAS.507.6108O}.

\begin{figure}
    \centering
    \includegraphics[width=1\linewidth]{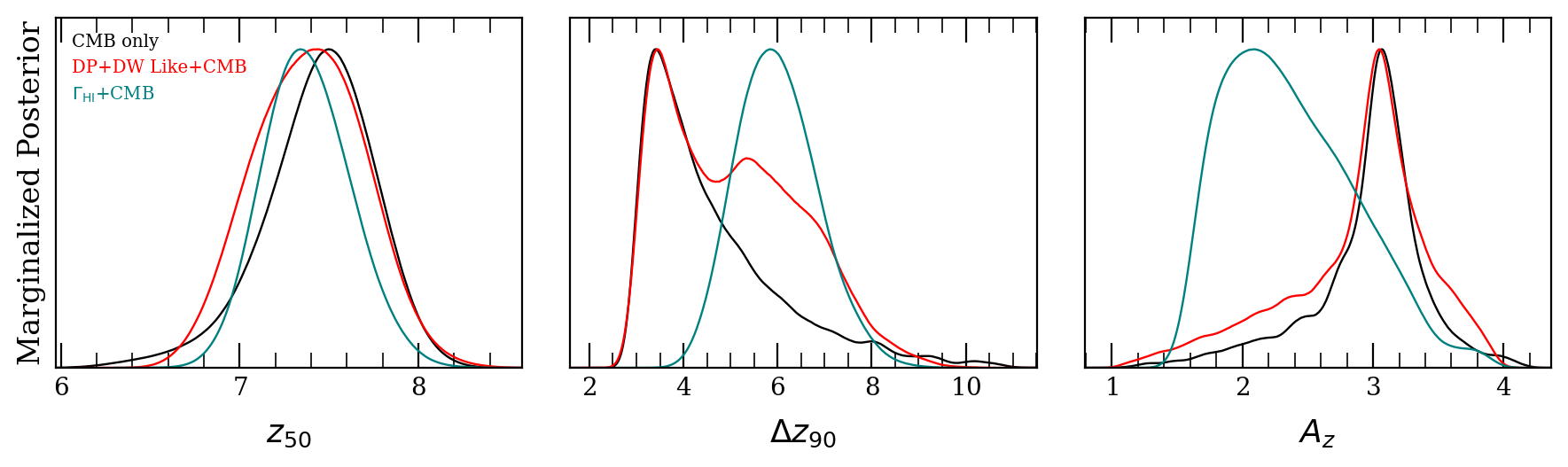}
    \caption{Constraints on $z_{50}$ (left), $\Delta z_{90}$ (center) and $A_z$ (right) in joint CMB-forest analyses, including Ly$\alpha$ forest measurements of the IGM photoionization rate (\textsc{$\Gamma_{\rm HI}$+CMB}).  Combining tight constraints on reionization's end (provided here by the $\Gamma_{\rm HI}$ measurements) with the CMB turns the latter from primarily a constraint on the midpoint of reionization into a more sensitive probe of the reionization history's shape. }
    \label{fig:DPLem_zmidDz_1d}
\end{figure}

Although the forest measurements directly constrain $\Gamma_{\rm HI}$ only at $z\lesssim6$, the semi-analytic model connecting $\Gamma_{\rm HI}$ to the emissivity and reionization history allows this information to propagate to higher redshifts. This can be seen in the middle two rows, which show the corresponding effect on the inferred reionization history. As we saw in the previous sections, the CMB combined with direct neutral fraction constraints still allows a broad range of histories with different durations and completion redshifts.  In contrast, the
\textsc{$\Gamma_{\rm HI}$+CMB} analysis localizes the final stages of reionization to $z\sim6-8$. Notably, these histories are also broadly consistent with the neutral-fraction constraints from Refs.~\cite{Zhu2024-aq,Davies2025-lf,Jin2023-bh}, even though those measurements are not included in the \textsc{$\Gamma_{\rm HI}$+CMB} likelihood.

The origin of this sensitivity can be understood from the dependence of the ionizing background on both the emissivity and the mean free path of ionizing photons. During reionization,
$\Gamma_{\rm HI}\propto\dot{n}_{\rm ion}\lambda_{\rm mfp}$, up to factors describing the ionizing radiation spectrum and column density distribution of absorbers. Near the end of reionization, the mean free path grows rapidly as the remaining neutral regions disappear and the distribution of ionizing absorbers evolves and the last islands of neutral hydrogen disappear. Within our model, a history that completes reionization too early therefore tends to produce a mean free path, and hence a photoionization rate, that is too large at fixed redshift. Conversely, completing reionization too late suppresses $\Gamma_{\rm HI}$ below the observed values because the ionizing background is not able to grow enough when neutral islands are still present. Matching the measured amplitude and redshift evolution of $\Gamma_{\rm HI}$ therefore restricts both the timing of the end of reionization and the evolution of the emissivity. Indeed, Ref.~\cite{Cain2026-op} found that shifting the end of reionization by only $\Delta z\sim0.5$ can change the predicted $\Gamma_{\rm HI}$ by more than the quoted $1\sigma$ observational uncertainties\footnote{They also noted an intrinsic modeling bias in their model, but they found that this is also smaller than the observational uncertainties.  }.  

The bottom row Figure \ref{fig:DPL_xHII_hist} highlights this connection. In the \textsc{CMB Only} and \textsc{DP+DW like+CMB} runs, the allowed $\Gamma_{\rm HI}$ spans a few orders of magnitude. Incorporating the $\Gamma_{\rm HI}$ constraints, as expected, produces a much narrower band that follows both the normalization and redshift evolution of the observed photoionization rate. The key result, however, is the corresponding reduction in the allowed emissivity and reionization histories. The CMB limits the amount of ionization at high redshift, whereas the Ly$\alpha$ forest anchors the ionizing background near the end of reionization; together, they constrain substantially more of the reionization history than either data set alone.

Figure~\ref{fig:DPLem_zmidDz_1d} shows the marginalized constraints on the midpoint and duration, which are also summarized in Table \ref{tab:parameters}.  The \textsc{CMB Only} and \textsc{DP+DWlike+CMB} posteriors both favor $z_{50}\simeq 7.5$ and relatively short reionization histories, although their $\Delta z_{90}$ posteriors retain extended tails toward larger $\Delta z_{90}$. The secondary peak in the \textsc{DP+DWlike+CMB} posterior arises from the same competition between the CMB and the lower limit on $x_{\rm HI}$ from Ref.~\cite{Zhu2024-aq} discussed in \S\ref{sec:parametric}. Although the posteriors for $\Delta z_{90}$ and $A_z$ in the \textsc{CMB Only} and \textsc{DP+DWlike+CMB} analyses appear to close on both sides, we will find below that the shapes of these are highly sensitive to the implicit priors, i.e. that the data are not very constraining on these quantities. We therefore quote only a 95\% upper limit for $\Delta z_{90}$ in Table~\ref{tab:parameters} for these analyses. 

Including the $\Gamma_{\rm HI}$ measurements has only a modest effect on the $z_{50}$ posterior, but shifts and substantially narrows the posterior for $\Delta z_{90}$ toward longer durations. The posterior for $A_z$ is shifted toward more symmetric reionization histories.  The origin of the longer duration can be understood from the inferred evolution of the ionized fraction. Matching the observed evolution of $\Gamma_{\rm HI}$ favors a later completion of reionization, with the IGM retaining a larger neutral fraction to lower redshift.  This delayed completion is compensated by an earlier start to reionization in order to preserve roughly the same $\tau_{\rm es}$, as required by the CMB constraints. Most of this additional high-redshift ionization occurs at relatively small $x_{\rm HII}$, allowing $z_{50}$ to remain broadly similar to the \textsc{CMB Only} and \textsc{DP+DWlike+CMB} results. The resulting histories in the \textsc{$\Gamma_{\rm HI}$+CMB} analysis are therefore more extended and symmetric.  The decreased asymmetry is explained by $z_{05}$ and $z_{95}$ increasing/decreasing by roughly similar amounts, which brings $A_z$ down.  In the \textsc{$\Gamma_{\rm HI}$+CMB} analysis (with fixed IGM opacity and cosmological parameters), we find 

\begin{equation}
z_{50}=7.4^{+0.2}_{-0.3}
\label{eq:GammaHI+CMB_A}
\end{equation}
\begin{equation}
\Delta z_{90}=6.0^{+0.8}_{-1.0}
\label{eq:GammaHI+CMB_B}
\end{equation}
\begin{equation}
A_z=2.4^{+0.3}_{-0.7}.
\label{eq:GammaHI+CMB_C}
\end{equation}
The central conclusion here is that combining tight constraints on reionization's end (provided here by the $\Gamma_{\rm HI}$ measurements) with the CMB turns the latter from primarily a constraint on the midpoint of reionization into a more sensitive probe of the reionization history's shape. This confirms the complementarity first identified by Ref.~\cite{2025arXiv250515899C} and extends that result beyond the optical-depth-level by using the full CMB temperature and polarization power-spectrum likelihoods and performing a Bayesian analysis using a combination of constraints. 

\begin{figure}
    \centering
    \includegraphics[width=0.95\linewidth]{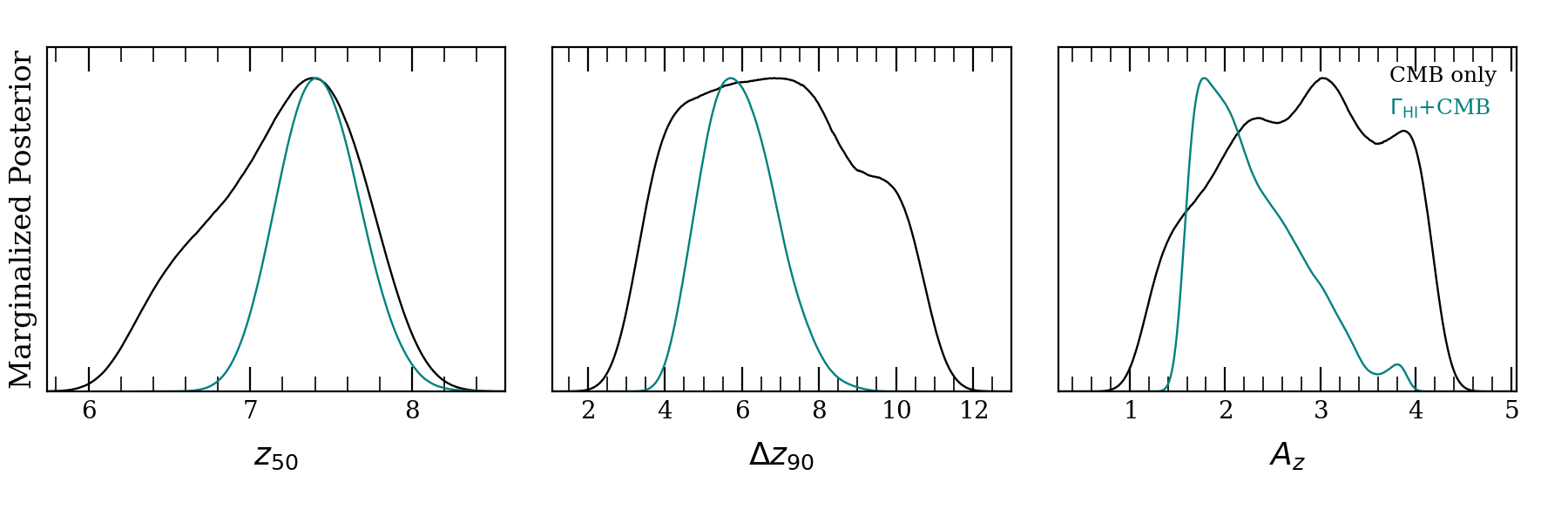}
    \caption{Similar to Figure \ref{fig:DPLem_zmidDz_1d}, except that the prior distribution in the space of $z_{50}$, $\Delta z_{50}$ and $A_z$ has been flattened using importance sampling. After flattening, the \textsc{CMB Only} posteriors widen significantly compared to those in Figure \ref{fig:DPLem_zmidDz_1d} (which adopt broad uniform priors in the double power-law emissivity parameters, eq. \ref{eqn:DPL}).  However, the \textsc{$\Gamma_{\rm HI}$+CMB} posteriors are very similar to those in Figure \ref{fig:DPLem_zmidDz_1d}, indicating that they are insensitive to the priors. This insensitivity owes to the increased constraining power from incorporating the $\Gamma_{\rm HI}$ measurements. }
    \label{fig:DPL_flattened}
\end{figure}

How sensitive are these constraints to the choice of priors?  We again address this question by flattening the prior distribution in $z_{50}$, $\Delta z_{90}$, and $A_z$, using the importance sampling methods described in the previous section.  The resulting posteriors for the \textsc{CMB Only} and \textsc{$\Gamma_{\rm HI}$+CMB} analyses are shown in Figure \ref{fig:DPL_flattened}. Comparing those results to Figure \ref{fig:DPLem_zmidDz_1d}, we find that the \textsc{CMB Only} posteriors for all three quantities widen significantly upon adopting a uniform prior. This effect owes to the physical constraints imposed by the equations of the semi-analytic model and the limited constraining power of the CMB likelihoods alone.  In our fiducial analysis, we assume broad uniform priors on the parameters of the double power-law emissivity, equation (\ref{eqn:DPL}). Propagating these priors through the physics equations of the semi-analytic model yields an implicit prior distribution for $z_{50}$, $\Delta z_{90}$, and $A_z$.  Flattening this distribution in post-processing effectively removes the physical constraints on these quantities, which accounts for the widening of the \textsc{CMB Only} posteriors in Figure \ref{fig:DPL_flattened}.  By contrast, the \textsc{$\Gamma_{\rm HI}$+CMB} posteriors are insensitive to the choice of priors, as evidenced by the similarity of those posteriors in Figures \ref{fig:DPLem_zmidDz_1d} and \ref{fig:DPL_flattened}.  After prior flattening, we find $z_{50}=7.4^{+0.2}_{-0.3}$, $\Delta z_{90}=5.9^{+0.8}_{-1.1}$, and $A_z=2.3^{+0.3}_{-0.7}$, which are very similar to (\ref{eq:GammaHI+CMB_A}), (\ref{eq:GammaHI+CMB_B}), and (\ref{eq:GammaHI+CMB_C}). The similarity owes to the increased constraining power afforded by the forest $\Gamma_{\rm HI}$ constraints.

We conclude this section with two important caveats.  First, we have used an approximate form of the $\Gamma_{\rm HI}$ likelihood for this analysis. Secondly, we have held the cosmological parameters, the clumping factor normalization, and its scaling with $\Gamma_{\rm HI}$ fixed. We have also adopted a simulation-calibrated form for the ionizing bubble size distribution in our reionization modeling (see eqs. \ref{eqn:lambda_N}, \ref{eqn:clumping} and surrounding text).   A more detailed analysis will require: (1) A more careful treatment of possible non-Gaussianities and covariances in the $\Gamma_{\rm HI}$ measurements; (2) Marginalizing over cosmology as well as the parameters describing ionizing photon absorption. The preceding analysis nevertheless illustrates the constraining power of the Ly$\alpha$ forest transmission when combined with the CMB.

\section{Conclusions}
\label{sec:conclusions}

We have investigated how Lyman-series forest measurements complement the CMB temperature and polarization anisotropies in constraining reionization. A CMB-only analysis permits a broad range of degenerate reionization histories, despite yielding tight constraints on the integrated Thomson optical depth. For instance, a relatively late midpoint can be compensated by an extended high-redshift tail, producing nearly the same optical depth and large-scale E-mode polarization signal.  Incorporating Lyman-series forest constraints on the end of reionization can alleviate the degeneracy and narrow the range of allowed histories. 

To explore the information content of different Lyman-series forest statistics, we have considered 3 forms of constraints: (1) Upper limits on the IGM neutral fraction at $4.9 \lesssim z \lesssim 6.9$ from dark pixel counts \cite{Davies2025-lf, Jin2023-bh}; (2) A lower limit on the neutral fraction at $z=5.8$ based on damping-wing-like features adjacent to Ly$\beta$ dark gaps in quasar spectrum stacks \cite{Zhu2024-aq}; (3) Measurements of the mean IGM photoionization rate at $4.9 \lesssim z \lesssim 6$ derived from Ly$\alpha$ forest transmission measurements \cite{Gaikwad2023-mm}. To explore model dependencies, we also compared 3 reionization modeling approaches.  We first considered a parametric model in which the reionization history is described by a Weibull function, characterized by a midpoint, duration, and asymmetry.  We also considered the more free-form \textsc{FlexKnots} approach.  Lastly, we considered a new physical model from Ref. \cite{Cain2026-op} based on the cosmological radiative transfer equation.

Our main conclusions can be summarized as follows:

\begin{itemize}
    \item The dark-pixel and damping-wing-like constraints provide complementary information on the reionization history.  Incorporating the dark pixel upper limits on $\langle x_{\rm HI}\rangle_V$ into a joint CMB-forest analysis imposes a lower-bound on the midpoint of reionization and favors shorter reionization durations relative to a CMB-only analysis.  On the other hand, incorporating the damping-wing-like lower limit on $\langle x_{\rm HI}\rangle_V$ favors later midpoints and more extended reionization histories. 
    
    \item Combining the dark pixel and damping-wing-like measurements with the CMB yields tighter constraints on reionization's midpoint and duration. In our \textsc{DP+DW like+CMB} analysis using the parametric reionization model, we find $z_{50}=7.5^{+0.5}_{-0.9}$ and $\Delta z_{90} < 11.5$ (95\% upper limit). For the same anaylsis using \textsc{FlexKnots}, we find $z_{50}=7.7^{+0.7}_{-0.8}$ and $\Delta z_{90} < 11.4$.   The asymmetry of the reionization history, however, is poorly constrained by these data.
    
    \item Neither the choice between our parametric and non-parametric descriptions, nor the choice among different forest data combinations, produces a significant shift in $\tau_{\rm es}$ or the other cosmological parameters considered here. Instead, the forest measurements primarily redistribute the optical depth in redshift. Constraints on the high-redshift ($z>15$) contribution to the optical depth are more sensitive to these choices. 
    
    \item Stronger constraints can be obtained by leveraging measurements of the forest transmission at $z \lesssim 6$. In this work, we have used measurements of the IGM photoionization rate, $\Gamma_{\rm HI}$, as  proxy for the mean transmission. Using the semi-analytic reionization model of Ref.~\cite{Cain2026-op}, we demonstrated that the $\Gamma_{\rm HI}$ measurements strongly restrict the allowed emissivity and reionization histories compared to the corresponding CMB-only analysis. In the \textsc{$\Gamma_{\rm HI}$+CMB} analysis (with fixed IGM opacity model and fixed cosmological parameters), we find $z_{50} = 7.4^{+0.2}_{-0.3}$, $\Delta z_{90}=6.0^{+0.8}_{-1.0}$, and $A_z = 2.4^{+0.3}_{-0.7}$. In qualitative agreement with Ref. \cite{2025arXiv250515899C}, our analysis confirms that the CMB becomes a more sensitive probe of the reionization history once the last phases of reionization are tightly constrained by forest measurements.

    \item{Our joint analysis with $\Gamma_{\rm HI}$ and CMB measurements favors models in which the ionizing emissivity of reionization sources begins declining after $z\approx 7-8$.  This evolution may owe to a decline in the escape fraction of ionizing photons and/or the onset of a feedback process regulating star formation. }

\end{itemize}

Taken together, our results show that combining the Lyman-series forest with CMB power spectra provides access to the chronology of reionization, rather than merely its integrated optical depth. Improvements in high-redshift forest measurements, their associated likelihood products, and large-scale CMB polarization measurements should therefore translate directly into stronger joint constraints on the full reionization history and on the evolution of the ionizing sources that drove it. A further challenge to this goal will be to develop computationally efficient frameworks that allow the Ly$\alpha$ forest transmission itself to be forward-modeled within a Bayesian MCMC inference pipeline.  At present, predicting the forest transmission with sufficient fidelity generally requires expensive radiative hydrodynamic simulations, making direct incorporation into inference pipelines impractical.  Our results provide additional motivation for developing fast and accurate forward-modeling techniques for the forest. 

\acknowledgments

The authors thank Shangqi Liu, Simeon Bird, and George Becker for helpful conversations.  AD used ChatGPT GPT-5.6 Sol to edit/sharpen some passages of text in this manuscript.  AD acknowledges support from grant NSF AST-2045600.  CC acknowledges support
from the Beus Center for Cosmic Foundations at Arizona State University. 
Computations were run using computational resources distributed under NSF ACCESS allocations TG-PHYS210041 and TG-PHY240332.  

\appendix

\section{Cosmological parameters and testing CMB likelihood choices}
\label{sec:PR3_vs_PR4}

\begin{figure}
    \centering
    \includegraphics[width=0.9\linewidth]{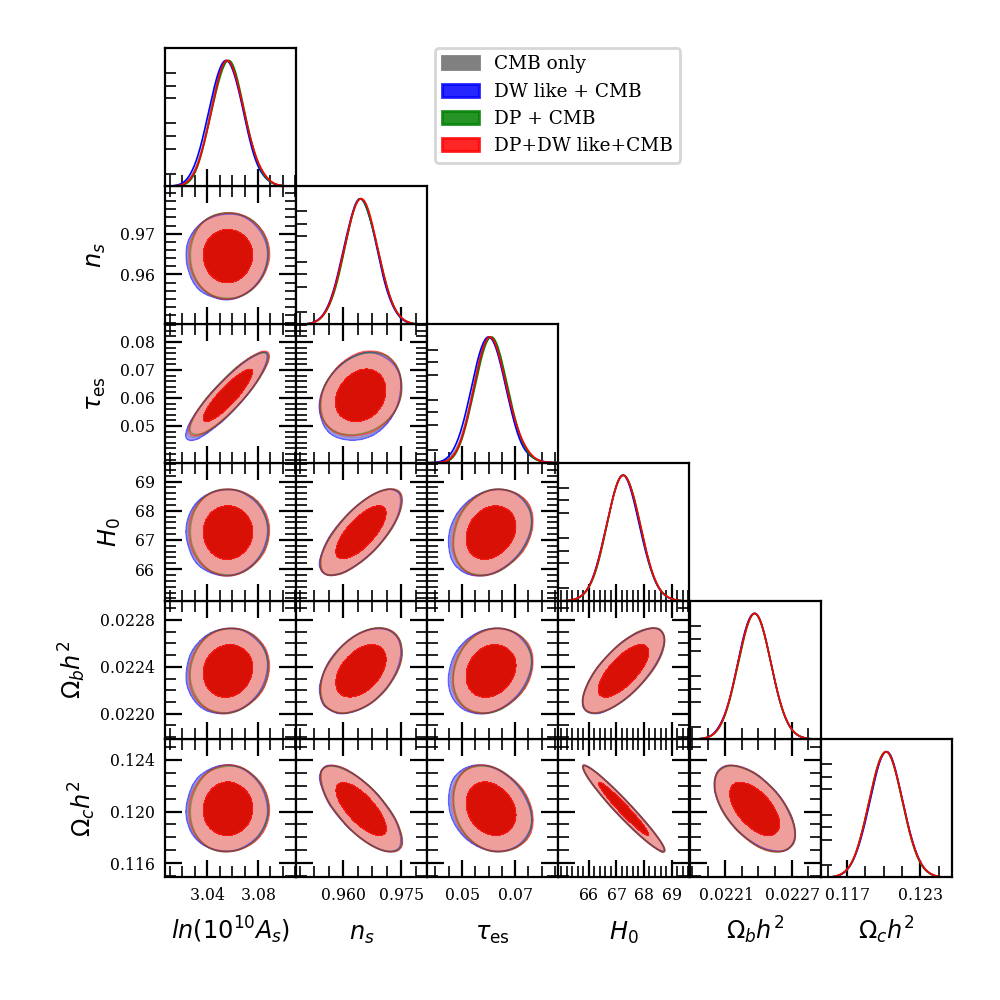}
    \caption{Corner plot for the 6 $\Lambda$CDM parameters obtained using the \textsc{FlexKnot} analysis of \S \ref{sec:FlexKnot}. We compare different dataset combinations as indicated in the plot legend.  The parameters are stable across different Lyman-series forest dataset combinations.  }
    \label{fig:Cosmo_corner}
\end{figure}

In Figure \ref{fig:Cosmo_corner}, we show a corner plot for the 6 $\Lambda$CDM parameters obtained using the \textsc{FlexKnot} analyses of \S \ref{sec:FlexKnot}.  The plot compares all of the different dataset combinations as indicated in the legend. We find that the cosmological parameters are stable to different combinations of the Lyman-series forest likelihoods. Although not shown in the figure, we have also compared these results to the cosmological parameters obtained using the parametric model in \S \ref{sec:parametric}, again finding no significant difference between them.  In Table \ref{tab:hillipop_cosmo_pars}, we provide a numerical comparison between our obtained values and the Planck PR3 and PR4 analyses (see discussion below). 

\begin{figure}
    \centering
    \includegraphics[width=1\linewidth]{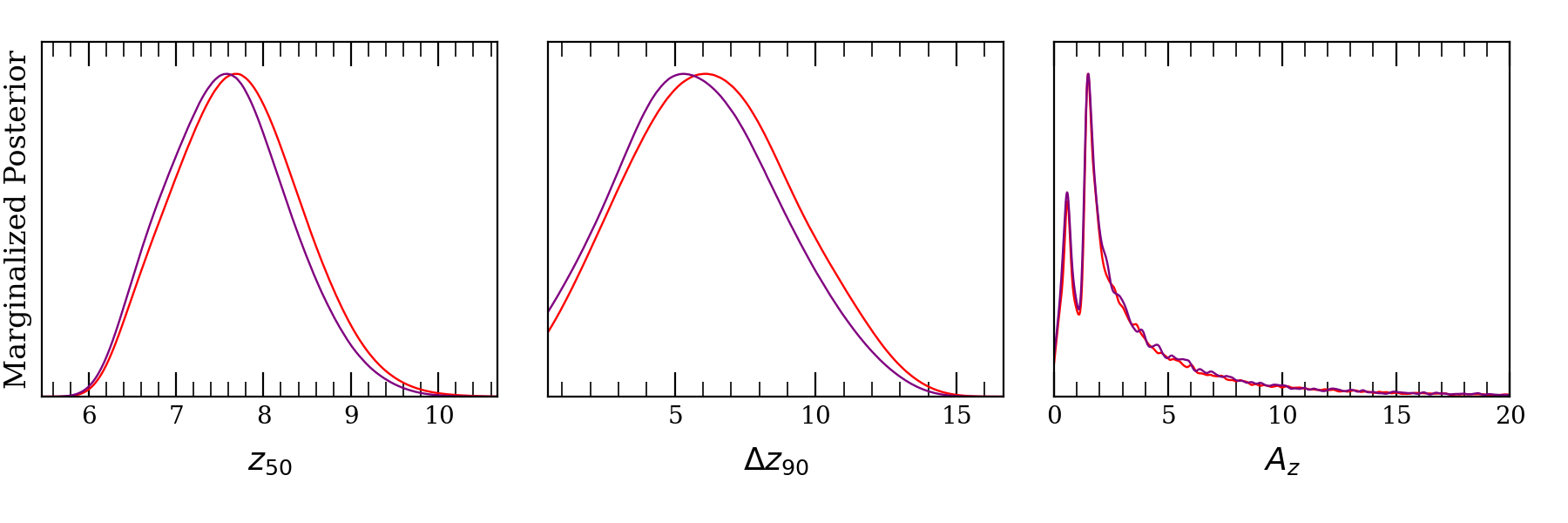}
    \caption{Same as Fig. \ref{fig:Deltaz_FlexKnot_corner}, but comparing our fiducial constraints (red) in the \textsc{DP+DWlike+CMB} case against those obtained when the \texttt{Plik$_{\rm lite}$} high-$\ell$ TTTEEE likelihood is replaced with \texttt{HiLLiPoP} (purple).  We see only small changes in the inferred reionization history and conclude that our results are insensitive to this choice.}
    \label{fig:hilli_compare}
\end{figure}

In this work we use the \texttt{Plik$_{\rm lite}$} and \texttt{Commander} likelihoods for high-$\ell$ TTTEEE and low-$\ell$ TT, respectively. Both are based on Planck PR3.  For low-$\ell$ EE, we use the more recent \texttt{Lollipop} likelihood (PR4).  Given changes to the data reduction pipeline between PR3 and PR4, in this section we also verify that the choice to mix PR3 and PR4 likelihoods does not significantly affect our main results. We perform a separate \textsc{DP+DWlike+CMB} analysis under our \textsc{FlexKnot} model where we replace \texttt{Plik$_{\rm lite}$} with \texttt{HiLLiPoP} as the high-$\ell$ TTTEEE likelihood. In this case, the likelihoods are fully up-to-date with PR4 since there were no updates to the \texttt{Commander} low-$\ell$ TT likelihood in the PR4 analysis. We find little difference with our fiducial analysis in the reionization history posteriors.  In Figure, \ref{fig:hilli_compare}, we compare our fiducial constraints (red) on $z_{50}$, $\Delta z_{90}$, and $A_z$, against the case with  \texttt{HiLLiPoP} (purple).

Table \ref{tab:hillipop_cosmo_pars} compares the cosmological parameters inferred from our fiducial analysis (4th column) and \texttt{HiLLiPoP} test (5th column) against Planck PR3 \cite{Planck18} (2nd column) and the PR4-based results of \cite{Tristram2024} (3rd column). Relative to the PR3 baseline, our fiducial  \texttt{Plik}-based results exhibit a mild shift toward larger values of $\tau_{\rm es}$, $A_s$, and $\sigma_8$, with each difference amounting to approximately $0.6-0.7\sigma$ when the quoted uncertainties are combined in quadrature. These shifts are not independent, but instead follow the $\tau_{\rm es}$--$A_s$--$\sigma_8$ degeneracy. The remaining cosmological parameters are effectively unchanged. In contrast, the \texttt{HiLLiPoP}-based results closely reproduce the PR4-based constraints from \cite{Tristram2024} across all parameters. Based on the results of this section, we conclude that our choice of CMB likelihoods does not affect the conclusions of our work.

\begin{table}
\resizebox{\textwidth}{!}{
    \centering
    \begin{tabular}{c|c|c|c|c}
         Parameter          & Planck PR3 \cite{Planck18}           & Ref. \cite{Tristram2024} (PR4-based)            & \textsc{DP+DWlike+CMB} (\textsc{Plik})             & \textsc{DP+DWlike+CMB} (\textsc{HiLLiPoP})         \\
         log$(10^{10} A_s)$ & $3.044\pm 0.014$      & $3.040\pm 0.014$      & $3.057 \pm 0.013$             & $3.041\pm0.014$               \\ 
         $n_s$              & $0.9649\pm0.0042$     & $0.9681 \pm 0.0039$   & $0.9646\pm0.0043$             & $0.9676\pm0.0040 $           \\
         $\tau_{\rm es}$    & $0.0544\pm0.0073$     & $0.0580 \pm 0.0062$   & $0.0610^{+0.0058}_{-0.0065}$  & $0.0594 \pm 0.0061$          \\
         $\Omega_bh^2$      & $0.02237\pm0.0015$    & $0.02226 \pm 0.00013$ & $0.02237\pm 0.00015$          & $0.02225\pm0.000126$          \\
         $\Omega_c h^2$     & $0.1200\pm0.0012$     & $0.1188 \pm 0.0012$   & $0.1202\pm0.0014$             & $0.1188 \pm 0.0012$           \\
         $100\theta_*$      & $1.04110\pm0.00031$   & $1.04108\pm 0.00026$  & $1.04109\pm0.00031$           & $1.0411\pm0.00025$            \\
         $H_0$              & $67.36\pm0.54$        & $67.64 \pm 0.52$      & $67.26\pm0.60$                & $67.64\pm0.53$                \\
         $\sigma_8$     & $0.8111\pm0.0060$         & $0.8070 \pm 0.0065$   & $0.8169 \pm 0.0067$           & $0.8073\pm0.0066$             \\
    \end{tabular}}
    \caption{Comparison of the cosmological parameters inferred from our fiducial analysis (4th column) and \texttt{HiLLiPoP} test (5th column) against Planck PR3 \cite{Planck18} (2nd column) and the PR4-based results of \cite{Tristram2024} (3rd column).} 
    \label{tab:hillipop_cosmo_pars} 
\end{table}

Ref.~\cite{Ilic2025-xj} performed a closely related CMB-only analysis using a \texttt{3-Flex} model, similar to our \textsc{FlexKnot} parameterization, obtaining $\tau_{\rm es}=0.0593^{+0.0057}_{-0.0062}$. Reweighting their posterior to impose a flat prior on $\tau_{\rm es}$ reduced this to $0.0579\pm0.0061$, demonstrating a modest effect due to choice of prior. The raw posterior, however, agrees remarkably well with our \texttt{HiLLiPoP}-based result, $\tau_{\rm es}=0.05936\pm0.0062$, as expected given the similar combination of CMB likelihoods.  The somewhat larger values of $\tau_{\rm es}$, $A_s$, and $\sigma_8$ in our fiducial \texttt{Plik}-based analysis therefore appear to arise primarily from the choice of CMB likelihoods. Consistent with this interpretation, replacing \texttt{LoLLiPoP} with \texttt{SimAll} in our CMB-only analysis gives $\tau_{\rm es}\simeq0.0572$ and $\ln(10^{10}A_s)\simeq3.050$, closer to the Planck 2018 constraints. Overall, the differences are small compared with the uncertainties.  Any such differences between our constraints on $\tau_{\rm es}$ and other works likely owe to both the implicit $\tau_{\rm es}$ priors and different choices in the CMB likelihoods.

\section{Volume-weighted versus mass-weighted reionization histories}
\label{sec:vol_ave}

\begin{figure}
    \centering
    \includegraphics[width=0.85\linewidth]{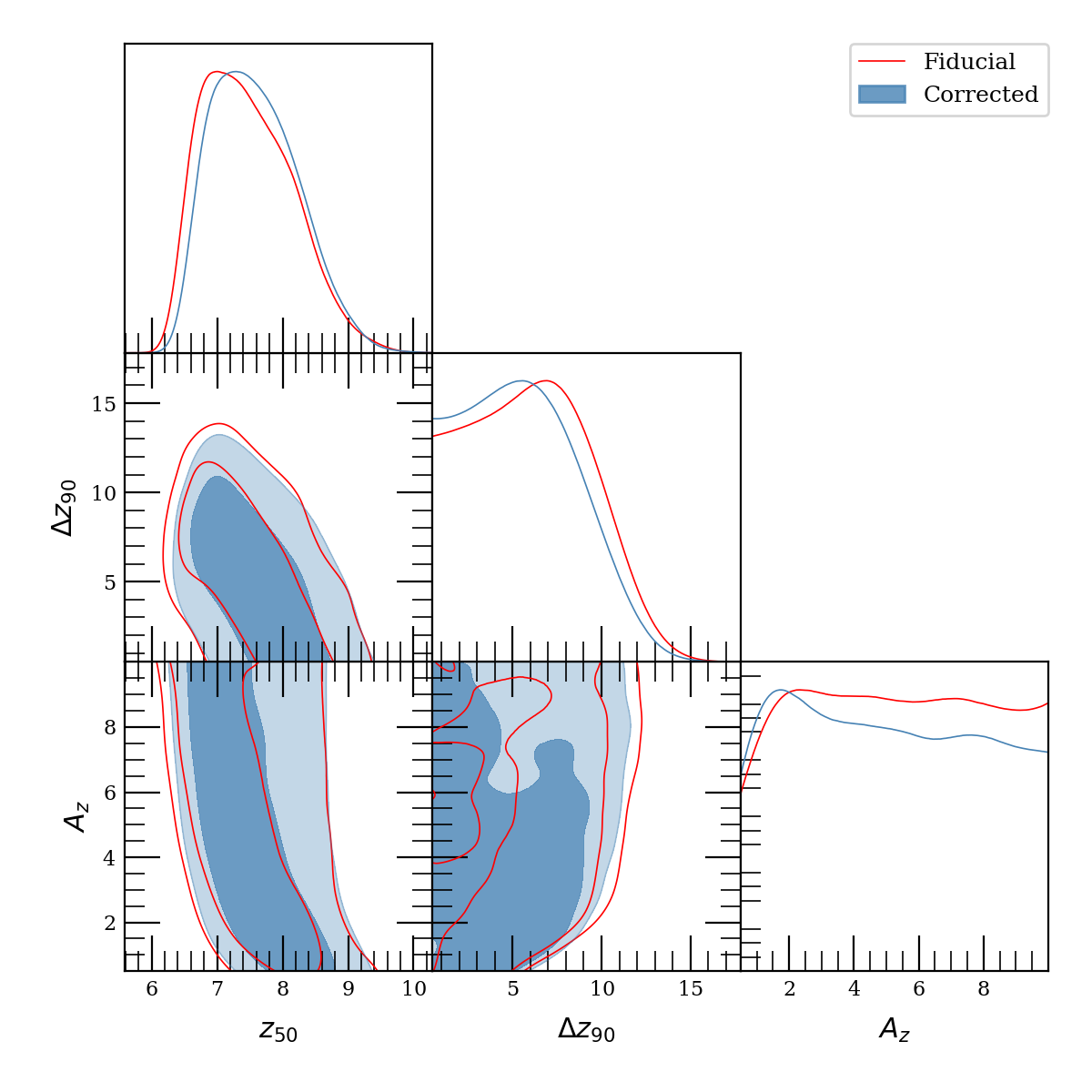}
    \caption{ Testing the effect of assuming $\langle x_{\rm HII} \rangle_V\approx \langle x_{\rm HII} \rangle_M$ in a joint analysis with Lyman-series forest constraints on $\langle x_{\rm HII} \rangle_V$ and CMB measurements.  The red curves and contours show results from our fiducial analysis using the \textsc{DP+DWlike+CMB} combination and parametric reionization model described in \S \ref{sec:parametric}.  The blue curves and contours correspond to a ``corrected'' analysis that uses a radiative transfer simulation of reionization to map between $\langle x_{\rm HII} \rangle_V$ and $\langle x_{\rm HII} \rangle_M$.  The corrected analysis is otherwise identical to the fiducial analysis. Accounting for the difference between volume- and mass-weighted ionized fractions has a weak effect on our results.}
    \label{fig:vol_Weibull}
\end{figure}

The dark pixel and damping-wing-like measurements described in \S \ref{sec:Forest_obs} place limits on the volume-weighted mean ionized fraction of the IGM, $\langle x_{\rm HII}\rangle_V$.  The CMB temperature and polarization anisotropies, however, are sensitive to the mass-weighted mean ionized fraction, $\langle x_{\rm HII} \rangle_M$.  In principle, this difference should be accounted for when combining constraints in a joint analysis. Broadly, simulations find that $\langle x_{\rm HII} \rangle_M$ leads $\langle x_{\rm HII}\rangle_V$ for most of reionization because over-dense regions are the first to reionize.  The difference can be as large as $\sim10-15\%$ near the midpoint of reionization, reducing to less than 5\% for $\langle x_{\rm HII}\rangle_M>90\%$.  Near the end of reionization, these differences are comparable to the $1\sigma$ uncertainties reported in the dark pixel and damping-wing-like measurements.  Applying both $\langle x_{\rm HII} \rangle_M$ and $\langle x_{\rm HII} \rangle_V$ in a joint analysis introduces an additional model dependence in order to convert one quantity into the other (see e.g. \cite{2026arXiv260518939S}). For simplicity, we have assumed $\langle x_{\rm HII} \rangle_M \approx \langle x_{\rm HII} \rangle_V$ throughout this paper. Here we test the effect of this assumption.   

We construct test set of likelihoods where we convert $\langle x_{\rm HII}\rangle_M \xrightarrow[]{} \langle x_{\rm HII}\rangle_V$ using a mapping obtained from a radiative transfer simulation of reionization. We construct an interpolation function using the Late Start/Late End simulation of Ref. \cite{Cain2024-wt}, chosen among the three simulations in that work because it maximizes differences between $\langle x_{\rm HII}\rangle_M$ and $\langle x_{\rm HII}\rangle_V$.   We use the interpolation function to convert the $\langle x_{\rm HII}\rangle_M$ from the CMB analysis into $\langle x_{\rm HII}\rangle_V$ before computing the likelihoods for the dark pixel and damping-wing-like constraints.  As an example, our conversion gives $\langle x_{\rm HII}\rangle_V\approx0.94$ for $\langle x_{\rm HII}\rangle_M=0.96$. We apply this model to our \textsc{DP+DWlike+CMB} analysis using the parametric reionization model of \S \ref{sec:parametric}. In Figure. \ref{fig:vol_Weibull}, we compare the posterior constraints from our fiducial analysis (red) against the ``corrected'' analysis (blue). Table \ref{tab:vol_Wei_param} summarizes the marginalized constraints on $z_{50}$, $\Delta_{90}$, and $A_z$.  Note that these quantities are still defined with respect to the mass-weighted reionization history.  The constraints change only modestly, with the corrected results preferring slightly higher $z_{50}$, shorter $\Delta z_{90}$, and slightly lower $A_z$. These differences, however, are well within the derived uncertainties. From these results we conclude that our main results and conclusions are not sensitive to the approximation $\langle x_{\rm HII}\rangle_V \approx \langle x_{\rm HII}\rangle_M$ adopted throughout this paper.

\begin{figure}
    \centering
    \includegraphics[width=0.92\linewidth]{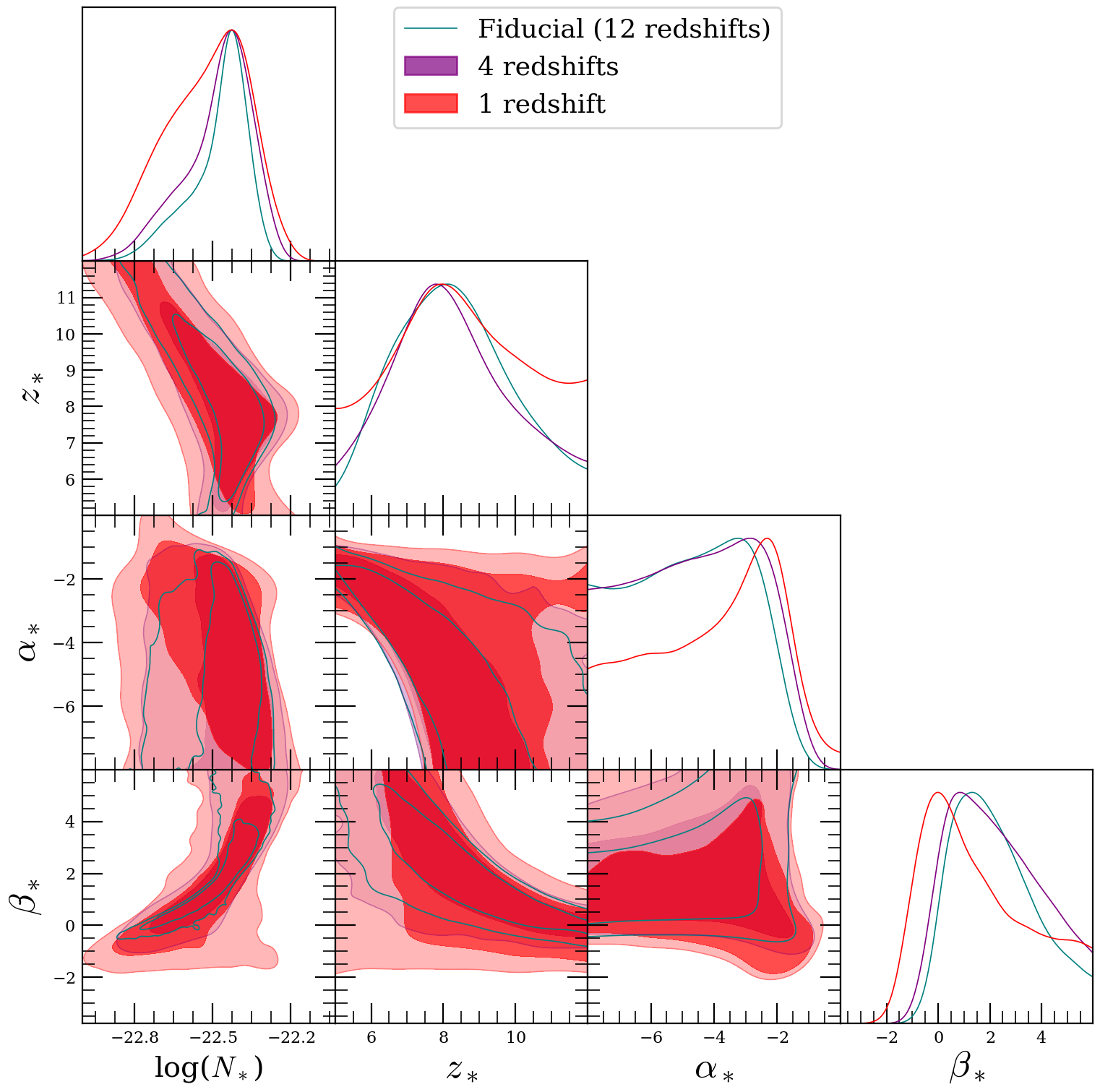}
    \caption{ Testing the effects of covariance between redshift bins in the \textsc{$\Gamma_{\rm HI}$+CMB} analysis of \S \ref{sec:results_Gamma}.  Our fiducial analysis uses $\Gamma_{\rm HI}$ measurements in 12 redshift bins.  We compare against a test run using just 4 of those bins with $\bar{z}=\{4.9, 5.2, 5.6, 6.0\}$.  Another test run uses just the $\bar{z}=5.5$ bin. These runs crudely mimic the effects of correlations among the 12 bins, with the latter bracketing the loss of constraining power that could result from strong correlations between redshift bins.}
    \label{fig:covar_corner}
\end{figure}

\begin{table}
    \centering
    \begin{tabular}{l|c|c|c}
         Analysis   & $z_{50}$              & $\Delta z_{90}$, (95\% UL)    \\ [0.4ex]
         \hline 
         Fiducial   & $7.5^{+0.5}_{-0.9}$   & $<11.5$                       \\ [0.4ex]
         Corrected  & $7.6^{+0.5}_{-0.8}$   & $<10.9$                       \\ [0.4ex]
    \end{tabular}
    \caption{Posterior constraints on Wiebull function parameters using the \textsc{DP+DWlike+CMB} data set.  Here we compare our fiducial analysis, which assumes $\langle x_{\rm HII} \rangle_V \approx \langle x_{\rm HII}\rangle_M$, to a ``corrected'' analysis that uses a radiative transfer simulation of reionization to derive a mapping between these quantities. Accounting for the difference between $\langle x_{\rm HII} \rangle_V$ and $\langle x_{\rm HII} \rangle_M$ in the joint CMB-forest analysis has a modest effect on the posterior constraints.}
    \label{tab:vol_Wei_param}
\end{table}

\section{Covariance in $\Gamma_{\rm HI}$ measurements}
\label{sec:Gam_like_test}

\begin{figure}
    \centering
    \includegraphics[width=0.95\linewidth]{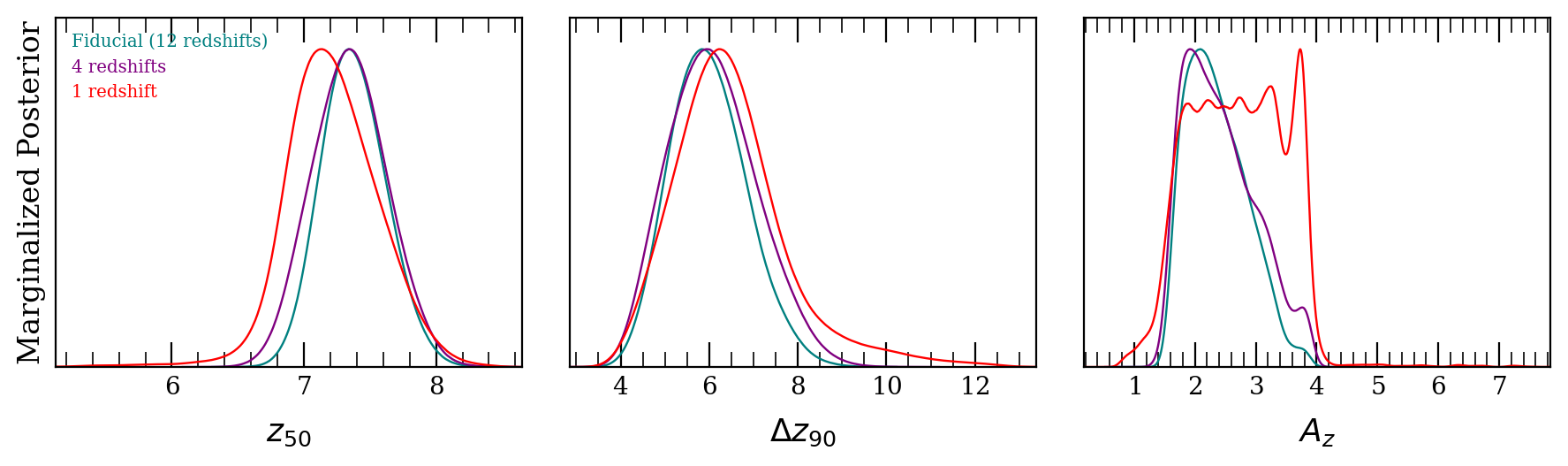}
    \caption{ Constraints on $z_{50}$, $\Delta z_{90}$ and $A_z$ for our fiducial (teal), 4-redshift (purple) and one-redshift (red) analyses.  The similarities, especially between the fiducial and 4-redshift results, suggest that our main conclusions in \S \ref{sec:results_Gamma} are robust to the neglect of covariances between the redshift bins.}
    \label{fig:covar_Dz}
\end{figure}

In \S \ref{sec:results_Gamma}, our approximate $\Gamma_{\rm HI}$ likelihood treats each redshift bin as independent.  In this section, we present a simple test that help gauge the impact of this approximation on our results.  Our fiducial \textsc{$\Gamma_{\rm HI}$+CMB} analysis in \S \ref{sec:results_Gamma} uses measurements at 12 redshifts. For our two test runs, we perform the same analysis but using only four (spaced by $\Delta z \approx 0.4$) and one of those measurements. Specifically, the four-redshift run uses the bins with $\bar{z}=\{4.9, 5.2, 5.6, 6.0\}$, while the one-redshift run uses the bin with $\bar{z}=5.5$.  These runs crudely mimic the effects of correlations among the 12 redshift bins, with the latter bracketing the loss of constraining power that could result from strong correlations between redshift bins.

In Figure \ref{fig:covar_corner} we compare constraints on emissivity parameters from the four-redshift and one-redshift analyses against our fiducial analysis.  The four-redshift results are quite similar to the fiducial results.  Even the extreme one-redshift analysis recovers the core qualitative features of the fiducial analysis, but with larger uncertainties. A noteworthy difference in the one-redshift analysis is the diminished preference for $\beta_*>0$ seen in the bottom-right panel of Figure \ref{fig:covar_corner}, i.e. a diminished preference for a declining $\dot{n}_{\rm em}$ near the end of reionization. Naturally, constraining $\Gamma_{\rm HI}$ at only one redshift severely limits information about the $z < z_*$ slope of $\dot{n}_{\rm em}$. In Figure \ref{fig:covar_Dz} we show the fully marginalized posteriors for $z_{50}$, $\Delta z_{90}$ $A_z$ .  Again, the four-redshift analysis gives results very similar to the fiducial case.  Notably, the posterior for $\Delta z_{90}$ remains broadly intact even in the one-redshift analysis.   

These simple tests suggest that our main conclusions in \S \ref{sec:results_Gamma} would be broadly the same in a full analysis that accounts for the covariance between $\Gamma_{\rm HI}$ redshift bins.  

\bibliographystyle{JHEP}
\bibliography{reference}

@string{mnras = "MNRAS"}

@ARTICLE{1992MNRAS.256P..43E,
   author = {{Efstathiou}, G.},
    title = "{Suppressing the formation of dwarf galaxies via photoionization}",
  journal = {\mnras},
     year = 1992,
    month = may,
   volume = 256,
    pages = {43P-47P},
   adsurl = {http://adsabs.harvard.edu/abs/1992MNRAS.256P..43E}
}

@ARTICLE{2021MNRAS.507.2405C,
       author = {{Chatterjee}, Atrideb and {Choudhury}, Tirthankar Roy and {Mitra}, Sourav},
        title = "{CosmoReionMC: a package for estimating cosmological and astrophysical parameters using CMB, Lyman-{\ensuremath{\alpha}} absorption, and global 21 cm data}",
      journal = {\mnras},
         year = 2021,
        month = oct,
       volume = {507},
       number = {2},
        pages = {2405-2422},
          doi = {10.1093/mnras/stab2316},
archivePrefix = {arXiv},
       eprint = {2101.11088},
 primaryClass = {astro-ph.CO},
       adsurl = {https://ui.adsabs.harvard.edu/abs/2021MNRAS.507.2405C}
}

@ARTICLE{becker15,
   author = {{Becker}, G.~D. and {Bolton}, J.~S. and {Madau}, P. and {Pettini}, M. and 
	{Ryan-Weber}, E.~V. and {Venemans}, B.~P.},
    title = "{Evidence of patchy hydrogen reionization from an extreme Ly{$\alpha$} trough below redshift six}",
  journal = {\mnras},
archivePrefix = "arXiv",
   eprint = {1407.4850},
     year = 2015,
    month = mar,
   volume = 447,
    pages = {3402-3419},
      doi = {10.1093/mnras/stu2646},
   adsurl = {http://adsabs.harvard.edu/abs/2015MNRAS.447.3402B}
}

@ARTICLE{2019arXiv190512640K,
       author = {{Keating}, Laura C. and {Weinberger}, Lewis H. and {Kulkarni}, Girish and
         {Haehnelt}, Martin G. and {Chardin}, Jonathan and {Aubert}, Dominique},
        title = "{Long troughs in the Lyman-{\ensuremath{\alpha}} forest below redshift 6 due to islands of neutral hydrogen}",
      journal = {\mnras},
         year = 2020,
        month = jan,
       volume = {491},
       number = {2},
        pages = {1736-1745},
          doi = {10.1093/mnras/stz3083},
archivePrefix = {arXiv},
       eprint = {1905.12640},
 primaryClass = {astro-ph.CO},
       adsurl = {https://ui.adsabs.harvard.edu/abs/2020MNRAS.491.1736K}
}

@ARTICLE{2018arXiv180704229E,
   author = {{Eilers}, A.-C. and {Davies}, F.~B. and {Hennawi}, J.~F.},
    title = "{The Opacity of the Intergalactic Medium Measured along Quasar Sightlines at $z\lesssim6$}",
  journal = {\apj},
archivePrefix = "arXiv",
   eprint = {1807.04229},
     year = 2018,
    month = sep,
   volume = 864,
      eid = {53},
    pages = {53},
      doi = {10.3847/1538-4357/aad4fd},
   adsurl = {http://adsabs.harvard.edu/abs/2018ApJ...864...53E}
}

@ARTICLE{1996ApJ...456..413N,
       author = {{Ng}, Ka Lok and {Ng}, Kin-Wang},
        title = "{Large-Angle Polarization and Anisotropy of the Cosmic Microwave Background Radiation and Reionization}",
      journal = {\apj},
         year = 1996,
        month = jan,
       volume = {456},
        pages = {413},
          doi = {10.1086/176665},
archivePrefix = {arXiv},
       eprint = {astro-ph/9412097},
 primaryClass = {astro-ph},
       adsurl = {https://ui.adsabs.harvard.edu/abs/1996ApJ...456..413N}
}

@ARTICLE{1997PhRvD..55.1822Z,
       author = {{Zaldarriaga}, Matias},
        title = "{Polarization of the microwave background in reionized models}",
      journal = {\prd},
         year = 1997,
        month = feb,
       volume = {55},
       number = {4},
        pages = {1822-1829},
          doi = {10.1103/PhysRevD.55.1822},
archivePrefix = {arXiv},
       eprint = {astro-ph/9608050},
 primaryClass = {astro-ph},
       adsurl = {https://ui.adsabs.harvard.edu/abs/1997PhRvD..55.1822Z}
}

@ARTICLE{2020A&A...641A...5P,
       author = {{Planck Collaboration} and {Aghanim}, N. and {Akrami}, Y. and {Ashdown}, M. and {Aumont}, J. and {Baccigalupi}, C. and {Ballardini}, M. and {Banday}, A.~J. and {Barreiro}, R.~B. and {Bartolo}, N. and {Basak}, S. and {Benabed}, K. and {Bernard}, J.-P. and {Bersanelli}, M. and {Bielewicz}, P. and {Bock}, J.~J. and {Bond}, J.~R. and {Borrill}, J. and {Bouchet}, F.~R. and {Boulanger}, F. and {Bucher}, M. and {Burigana}, C. and {Butler}, R.~C. and {Calabrese}, E. and {Cardoso}, J.-F. and {Carron}, J. and {Casaponsa}, B. and {Challinor}, A. and {Chiang}, H.~C. and {Colombo}, L.~P.~L. and {Combet}, C. and {Crill}, B.~P. and {Cuttaia}, F. and {de Bernardis}, P. and {de Rosa}, A. and {de Zotti}, G. and {Delabrouille}, J. and {Delouis}, J.-M. and {Di Valentino}, E. and {Diego}, J.~M. and {Dor{\'e}}, O. and {Douspis}, M. and {Ducout}, A. and {Dupac}, X. and {Dusini}, S. and {Efstathiou}, G. and {Elsner}, F. and {En{\ss}lin}, T.~A. and {Eriksen}, H.~K. and {Fantaye}, Y. and {Fernandez-Cobos}, R. and {Finelli}, F. and {Frailis}, M. and {Fraisse}, A.~A. and {Franceschi}, E. and {Frolov}, A. and {Galeotta}, S. and {Galli}, S. and {Ganga}, K. and {G{\'e}nova-Santos}, R.~T. and {Gerbino}, M. and {Ghosh}, T. and {Giraud-H{\'e}raud}, Y. and {Gonz{\'a}lez-Nuevo}, J. and {G{\'o}rski}, K.~M. and {Gratton}, S. and {Gruppuso}, A. and {Gudmundsson}, J.~E. and {Hamann}, J. and {Handley}, W. and {Hansen}, F.~K. and {Herranz}, D. and {Hivon}, E. and {Huang}, Z. and {Jaffe}, A.~H. and {Jones}, W.~C. and {Keih{\"a}nen}, E. and {Keskitalo}, R. and {Kiiveri}, K. and {Kim}, J. and {Kisner}, T.~S. and {Krachmalnicoff}, N. and {Kunz}, M. and {Kurki-Suonio}, H. and {Lagache}, G. and {Lamarre}, J.-M. and {Lasenby}, A. and {Lattanzi}, M. and {Lawrence}, C.~R. and {Le Jeune}, M. and {Levrier}, F. and {Lewis}, A. and {Liguori}, M. and {Lilje}, P.~B. and {Lilley}, M. and {Lindholm}, V. and {L{\'o}pez-Caniego}, M. and {Lubin}, P.~M. and {Ma}, Y.-Z. and {Mac{\'\i}as-P{\'e}rez}, J.~F. and {Maggio}, G. and {Maino}, D. and {Mandolesi}, N. and {Mangilli}, A. and {Marcos-Caballero}, A. and {Maris}, M. and {Martin}, P.~G. and {Mart{\'\i}nez-Gonz{\'a}lez}, E. and {Matarrese}, S. and {Mauri}, N. and {McEwen}, J.~D. and {Meinhold}, P.~R. and {Melchiorri}, A. and {Mennella}, A. and {Migliaccio}, M. and {Millea}, M. and {Miville-Desch{\^e}nes}, M.-A. and {Molinari}, D. and {Moneti}, A. and {Montier}, L. and {Morgante}, G. and {Moss}, A. and {Natoli}, P. and {N{\o}rgaard-Nielsen}, H.~U. and {Pagano}, L. and {Paoletti}, D. and {Partridge}, B. and {Patanchon}, G. and {Peiris}, H.~V. and {Perrotta}, F. and {Pettorino}, V. and {Piacentini}, F. and {Polenta}, G. and {Puget}, J.-L. and {Rachen}, J.~P. and {Reinecke}, M. and {Remazeilles}, M. and {Renzi}, A. and {Rocha}, G. and {Rosset}, C. and {Roudier}, G. and {Rubi{\~n}o-Mart{\'\i}n}, J.~A. and {Ruiz-Granados}, B. and {Salvati}, L. and {Sandri}, M. and {Savelainen}, M. and {Scott}, D. and {Shellard}, E.~P.~S. and {Sirignano}, C. and {Sirri}, G. and {Spencer}, L.~D. and {Sunyaev}, R. and {Suur-Uski}, A.-S. and {Tauber}, J.~A. and {Tavagnacco}, D. and {Tenti}, M. and {Toffolatti}, L. and {Tomasi}, M. and {Trombetti}, T. and {Valiviita}, J. and {Van Tent}, B. and {Vielva}, P. and {Villa}, F. and {Vittorio}, N. and {Wandelt}, B.~D. and {Wehus}, I.~K. and {Zacchei}, A. and {Zonca}, A.},
        title = "{Planck 2018 results. V. CMB power spectra and likelihoods}",
      journal = {\aap},
         year = 2020,
        month = sep,
       volume = {641},
          eid = {A5},
        pages = {A5},
          doi = {10.1051/0004-6361/201936386},
archivePrefix = {arXiv},
       eprint = {1907.12875},
 primaryClass = {astro-ph.CO},
       adsurl = {https://ui.adsabs.harvard.edu/abs/2020A&A...641A...5P}
}

@ARTICLE{2022PhRvD.105h3524T,
       author = {{Tristram}, M. and {Banday}, A.~J. and {G{\'o}rski}, K.~M. and {Keskitalo}, R. and {Lawrence}, C.~R. and {Andersen}, K.~J. and {Barreiro}, R.~B. and {Borrill}, J. and {Colombo}, L.~P.~L. and {Eriksen}, H.~K. and {Fernandez-Cobos}, R. and {Kisner}, T.~S. and {Mart{\'\i}nez-Gonz{\'a}lez}, E. and {Partridge}, B. and {Scott}, D. and {Svalheim}, T.~L. and {Wehus}, I.~K.},
        title = "{Improved limits on the tensor-to-scalar ratio using BICEP and Planck data}",
      journal = {\prd},
         year = 2022,
        month = apr,
       volume = {105},
       number = {8},
          eid = {083524},
        pages = {083524},
          doi = {10.1103/PhysRevD.105.083524},
archivePrefix = {arXiv},
       eprint = {2112.07961},
 primaryClass = {astro-ph.CO},
       adsurl = {https://ui.adsabs.harvard.edu/abs/2022PhRvD.105h3524T}
}

@ARTICLE{2021A&A...647A.128T,
       author = {{Tristram}, M. and {Banday}, A.~J. and {G{\'o}rski}, K.~M. and {Keskitalo}, R. and {Lawrence}, C.~R. and {Andersen}, K.~J. and {Barreiro}, R.~B. and {Borrill}, J. and {Eriksen}, H.~K. and {Fernandez-Cobos}, R. and {Kisner}, T.~S. and {Mart{\'\i}nez-Gonz{\'a}lez}, E. and {Partridge}, B. and {Scott}, D. and {Svalheim}, T.~L. and {Thommesen}, H. and {Wehus}, I.~K.},
        title = "{Planck constraints on the tensor-to-scalar ratio}",
      journal = {\aap},
         year = 2021,
        month = mar,
       volume = {647},
          eid = {A128},
        pages = {A128},
          doi = {10.1051/0004-6361/202039585},
archivePrefix = {arXiv},
       eprint = {2010.01139},
 primaryClass = {astro-ph.CO},
       adsurl = {https://ui.adsabs.harvard.edu/abs/2021A&A...647A.128T}
}

@ARTICLE{2011arXiv1104.2932L,
  title         = "The Cosmic Linear Anisotropy Solving System ({CLASS}) {I}:
                   Overview",
  author        = "Lesgourgues, Julien",
  journal       = "arXiv [astro-ph.IM]",
  pages         = "arXiv:1104.2932",
  month         =  apr,
  year          =  2011,
  archivePrefix = "arXiv",
  primaryClass  = "astro-ph.IM"
}

@ARTICLE{2000ApJ...538..473L,
       author = {{Lewis}, Antony and {Challinor}, Anthony and {Lasenby}, Anthony},
        title = "{Efficient Computation of Cosmic Microwave Background Anisotropies in Closed Friedmann-Robertson-Walker Models}",
      journal = {\apj},
         year = 2000,
        month = aug,
       volume = {538},
       number = {2},
        pages = {473-476},
          doi = {10.1086/309179},
archivePrefix = {arXiv},
       eprint = {astro-ph/9911177},
 primaryClass = {astro-ph},
       adsurl = {https://ui.adsabs.harvard.edu/abs/2000ApJ...538..473L}
}

@ARTICLE{2021JCAP...05..057T,
       author = {{Torrado}, Jes{\'u}s and {Lewis}, Antony},
        title = "{Cobaya: code for Bayesian analysis of hierarchical physical models}",
      journal = {\jcap},
         year = 2021,
        month = may,
       volume = {2021},
       number = {5},
          eid = {057},
        pages = {057},
          doi = {10.1088/1475-7516/2021/05/057},
archivePrefix = {arXiv},
       eprint = {2005.05290},
 primaryClass = {astro-ph.IM},
       adsurl = {https://ui.adsabs.harvard.edu/abs/2021JCAP...05..057T}
}

@ARTICLE{2025PASA...42...49Q,
  title     = "Percent-level timing of reionisation: Self-consistent,
               implicit-likelihood inference from {XQR}-30+ Ly$\alpha$ forest data",
  author    = "Qin, Yuxiang and Mesinger, Andrei and Prelogović, David and
               Becker, George and Bischetti, Manuela and Bosman, Sarah and
               Davies, Frederick and D'Odorico, Valentina and Gaikwad, Prakash
               and Haehnelt, Martin and Keating, Laura and Lai, Samuel and
               Ryan-Weber, Emma and Satyavolu, Sindhu and Walter, Fabian and
               Zhu, Yongda",
  journal   = "Publ. Astron. Soc. Aust.",
  publisher = "Cambridge University Press (CUP)",
  volume    =  42,
  number    = "e049",
  pages     = "e049",
  month     =  jan,
  year      =  2025,
  language  = "en"
}

@ARTICLE{2021PhRvD.104l3549P,
       author = {{Paoletti}, Daniela and {Hazra}, Dhiraj Kumar and {Finelli}, Fabio and {Smoot}, George F.},
        title = "{Dark twilight joined with the light of dawn to unveil the reionization history}",
      journal = {\prd},
         year = 2021,
        month = dec,
       volume = {104},
       number = {12},
          eid = {123549},
        pages = {123549},
          doi = {10.1103/PhysRevD.104.123549},
archivePrefix = {arXiv},
       eprint = {2107.10693},
 primaryClass = {astro-ph.CO},
       adsurl = {https://ui.adsabs.harvard.edu/abs/2021PhRvD.104l3549P}
}

@ARTICLE{2020PhRvL.125g1301H,
       author = {{Hazra}, Dhiraj Kumar and {Paoletti}, Daniela and {Finelli}, Fabio and {Smoot}, George F.},
        title = "{Joining Bits and Pieces of Reionization History}",
      journal = {\prl},
         year = 2020,
        month = aug,
       volume = {125},
       number = {7},
          eid = {071301},
        pages = {071301},
          doi = {10.1103/PhysRevLett.125.071301},
archivePrefix = {arXiv},
       eprint = {1904.01547},
 primaryClass = {astro-ph.CO},
       adsurl = {https://ui.adsabs.harvard.edu/abs/2020PhRvL.125g1301H}
}

@ARTICLE{2025PhRvD.111d3532P,
       author = {{Paoletti}, Daniela and {Hazra}, Dhiraj Kumar and {Finelli}, Fabio and {Smoot}, George F.},
        title = "{Evidence for asymmetric reionization histories from a joint analysis of the cosmic microwave background and astrophysical data}",
      journal = {\prd},
         year = 2025,
        month = feb,
       volume = {111},
       number = {4},
          eid = {043532},
        pages = {043532},
          doi = {10.1103/PhysRevD.111.043532},
archivePrefix = {arXiv},
       eprint = {2405.09506},
 primaryClass = {astro-ph.CO},
       adsurl = {https://ui.adsabs.harvard.edu/abs/2025PhRvD.111d3532P}
}

@ARTICLE{2024MNRAS.530.5209R,
       author = {{Roth}, Joshua T. and {D'Aloisio}, Anson and {Cain}, Christopher and {Wilson}, Bayu and {Zhu}, Yongda and {Becker}, George D.},
        title = "{The effect of reionization on direct measurements of the mean free path}",
      journal = {\mnras},
         year = 2024,
        month = jun,
       volume = {530},
       number = {4},
        pages = {5209-5219},
          doi = {10.1093/mnras/stae1194},
archivePrefix = {arXiv},
       eprint = {2311.06348},
 primaryClass = {astro-ph.CO},
       adsurl = {https://ui.adsabs.harvard.edu/abs/2024MNRAS.530.5209R}
}

@ARTICLE{2024MNRAS.533..676S,
       author = {{Satyavolu}, Sindhu and {Kulkarni}, Girish and {Keating}, Laura C. and {Haehnelt}, Martin G.},
        title = "{Robustness of direct measurements of the mean free path of ionizing photons in the epoch of reionization}",
      journal = {\mnras},
         year = 2024,
        month = sep,
       volume = {533},
       number = {1},
        pages = {676-686},
          doi = {10.1093/mnras/stae1717},
archivePrefix = {arXiv},
       eprint = {2311.06344},
 primaryClass = {astro-ph.CO},
       adsurl = {https://ui.adsabs.harvard.edu/abs/2024MNRAS.533..676S}
}

@ARTICLE{2025arXiv250515899C,
  title     = "The cosmic microwave background optical depth constrains the
               duration of reionization",
  author    = "Cain, Christopher and Van Engelen, Alexander and Croker, Kevin S
               and Kramer, Darby and D'Aloisio, Anson and Lopez, Garett",
  journal   = "Astrophys. J. Lett.",
  publisher = "American Astronomical Society",
  volume    =  987,
  number    =  2,
  pages     = "L29",
  month     =  jul,
  year      =  2025,
  language  = "en"
}

@ARTICLE{2021MNRAS.507.6108O,
       author = {{Ocvirk}, Pierre and {Lewis}, Joseph S.~W. and {Gillet}, Nicolas and {Chardin}, Jonathan and {Aubert}, Dominique and {Deparis}, Nicolas and {Th{\'e}lie}, {\'E}milie},
        title = "{Lyman-alpha opacities at z = 4-6 require low mass, radiatively-suppressed galaxies to drive cosmic reionization}",
      journal = {\mnras},
         year = 2021,
        month = nov,
       volume = {507},
       number = {4},
        pages = {6108-6117},
          doi = {10.1093/mnras/stab2502},
archivePrefix = {arXiv},
       eprint = {2105.01663},
 primaryClass = {astro-ph.CO},
       adsurl = {https://ui.adsabs.harvard.edu/abs/2021MNRAS.507.6108O}
}

@ARTICLE{2024MNRAS.531.1951C,
       author = {{Cain}, Christopher and {D'Aloisio}, Anson and {Lopez}, Garett and {Gangolli}, Nakul and {Roth}, Joshua T.},
        title = "{On the rise and fall of galactic ionizing output at the end of reionization}",
      journal = {\mnras},
         year = 2024,
        month = jun,
       volume = {531},
       number = {1},
        pages = {1951-1970},
          doi = {10.1093/mnras/stae1223},
archivePrefix = {arXiv},
       eprint = {2311.13638},
 primaryClass = {astro-ph.CO},
       adsurl = {https://ui.adsabs.harvard.edu/abs/2024MNRAS.531.1951C}
}

@ARTICLE{Tristram2024,
       author = {{Tristram}, M. and {Banday}, A.~J. and {Douspis}, M. and {Garrido}, X. and {G{\'o}rski}, K.~M. and {Henrot-Versill{\'e}}, S. and {Hergt}, L.~T. and {Ili{\'c}}, S. and {Keskitalo}, R. and {Lagache}, G. and {Lawrence}, C.~R. and {Partridge}, B. and {Scott}, D.},
        title = "{Cosmological parameters derived from the final Planck data release (PR4)}",
      journal = {\aap},
         year = 2024,
        month = feb,
       volume = {682},
          eid = {A37},
        pages = {A37},
          doi = {10.1051/0004-6361/202348015},
archivePrefix = {arXiv},
       eprint = {2309.10034},
 primaryClass = {astro-ph.CO},
       adsurl = {https://ui.adsabs.harvard.edu/abs/2024A&A...682A..37T}
}

@ARTICLE{2007MNRAS.376..534I,
       author = {{Iliev}, Ilian T. and {Mellema}, Garrelt and {Shapiro}, Paul R. and {Pen}, Ue-Li},
        title = "{Self-regulated reionization}",
      journal = {\mnras},
         year = 2007,
        month = apr,
       volume = {376},
       number = {2},
        pages = {534-548},
          doi = {10.1111/j.1365-2966.2007.11482.x},
archivePrefix = {arXiv},
       eprint = {astro-ph/0607517},
 primaryClass = {astro-ph},
       adsurl = {https://ui.adsabs.harvard.edu/abs/2007MNRAS.376..534I}
}

@ARTICLE{Reichardt2020,
  title     = "An improved measurement of the secondary cosmic microwave
               background anisotropies from the {SPT}-{SZ} + {SPTpol} surveys",
  author    = "Reichardt, C L and Patil, S and Ade, P A R and Anderson, A J and
               Austermann, J E and Avva, J S and Baxter, E and Beall, J A and
               Bender, A N and Benson, B A and Bianchini, F and Bleem, L E and
               Carlstrom, J E and Chang, C L and Chaubal, P and Chiang, H C and
               Chou, T L and Citron, R and Moran, C Corbett and Crawford, T M
               and Crites, A T and Haan, T de and Dobbs, M A and Everett, W and
               Gallicchio, J and George, E M and Gilbert, A and Gupta, N and
               Halverson, N W and Harrington, N and Henning, J W and Hilton, G C
               and Holder, G P and Holzapfel, W L and Hrubes, J D and Huang, N
               and Hubmayr, J and Irwin, K D and Knox, L and Lee, A T and Li, D
               and Lowitz, A and Luong-Van, D and McMahon, J J and Mehl, J and
               Meyer, S S and Millea, M and Mocanu, L M and Mohr, J J and
               Montgomery, J and Nadolski, A and Natoli, T and Nibarger, J P and
               Noble, G and Novosad, V and Omori, Y and Padin, S and Pryke, C
               and Ruhl, J E and Saliwanchik, B R and Sayre, J T and Schaffer, K
               K and Shirokoff, E and Sievers, C and Smecher, G and Spieler, H G
               and Staniszewski, Z and Stark, A A and Tucker, C and Vanderlinde,
               K and Veach, T and Vieira, J D and Wang, G and Whitehorn, N and
               Williamson, R and Wu, W L K and Yefremenko, V",
  journal   = "Astrophys. J.",
  publisher = "American Astronomical Society",
  volume    =  908,
  number    =  2,
  pages     =  199,
  month     =  feb,
  year      =  2021,
  language  = "en"
}

@ARTICLE{Beringue2025-ka,
  title     = "The Atacama Cosmology Telescope: {DR6} power spectrum foreground
               model and validation",
  author    = "Beringue, Benjamin and Surrao, Kristen M and Hill, J Colin and
               Atkins, Zachary and Battaglia, Nicholas and Bolliet, Boris and
               Calabrese, Erminia and Choi, Steve K and Clark, Susan E and
               Duivenvoorden, Adriaan J and Dunkley, Jo and Giardiello, Serena
               and Goldstein, Samuel and Hensley, Brandon S and Hložek, Renée
               and Jense, Hidde T and Kramer, Darby and La Posta, Adrien and
               Louis, Thibaut and Mehta, Yogesh and Moodley, Kavilan and Naess,
               Sigurd and Partridge, Bruce and Qu, Frank J and Ried Guachalla,
               Bernardita and Sehgal, Neelima and Sifón, Cristóbal and Staggs,
               Suzanne T and Trac, Hy and Van Engelen, Alexander and Wollack,
               Edward J",
  journal   = "J. Cosmol. Astropart. Phys.",
  publisher = "IOP Publishing",
  volume    =  2025,
  number    =  10,
  pages     =  082,
  month     =  oct,
  year      =  2025
}

@ARTICLE{2016MNRAS.460.1328D,
       author = {{Davies}, Frederick B. and {Furlanetto}, Steven R.},
        title = "{Large fluctuations in the hydrogen-ionizing background and mean free path following the epoch of reionization}",
      journal = {\mnras},
         year = 2016,
        month = aug,
       volume = {460},
       number = {2},
        pages = {1328-1339},
          doi = {10.1093/mnras/stw931},
archivePrefix = {arXiv},
       eprint = {1509.07131},
 primaryClass = {astro-ph.CO},
       adsurl = {https://ui.adsabs.harvard.edu/abs/2016MNRAS.460.1328D}
}

@ARTICLE{2024arXiv240508885B,
  title     = "Damping Wing absorption associated with a giant Ly$\alpha$ trough at $z < 6$: direct evidence for late-ending reionization",
  author    = "Becker, George D and Bolton, James S and Zhu, Yongda and Hashemi,
               Seyedazim",
  journal   = "Mon. Not. R. Astron. Soc.",
  publisher = "Oxford University Press (OUP)",
  volume    =  533,
  number    =  2,
  pages     = "1525--1540",
  month     =  aug,
  year      =  2024,
  language  = "en"
}

@ARTICLE{2024arXiv240512273S,
  title     = "Damping wings in the Lyman-\textit{$\alpha$} forest: A model-independent
               measurement of the neutral fraction at 5.4 $<$ \textit{z} $<$ 6.1",
  author    = "Spina, Benedetta and Bosman, Sarah E I and Davies, Frederick B
               and Gaikwad, Prakash and Zhu, Yongda",
  journal   = "Astron. Astrophys.",
  publisher = "EDP Sciences",
  volume    =  688,
  pages     = "L26",
  month     =  aug,
  year      =  2024,
  language  = "en"
}

@ARTICLE{2021MNRAS.506.2390Q,
       author = {{Qin}, Yuxiang and {Mesinger}, Andrei and {Bosman}, Sarah E.~I. and {Viel}, Matteo},
        title = "{Reionization and galaxy inference from the high-redshift Ly {\ensuremath{\alpha}} forest}",
      journal = {\mnras},
         year = 2021,
        month = sep,
       volume = {506},
       number = {2},
        pages = {2390-2407},
          doi = {10.1093/mnras/stab1833},
archivePrefix = {arXiv},
       eprint = {2101.09033},
 primaryClass = {astro-ph.CO},
       adsurl = {https://ui.adsabs.harvard.edu/abs/2021MNRAS.506.2390Q}
}

@ARTICLE{2024ApJ...965..134D,
       author = {{Davies}, Frederick B. and {Bosman}, Sarah E.~I. and {Gaikwad}, Prakash and {Nasir}, Fahad and {Hennawi}, Joseph F. and {Becker}, George D. and {Haehnelt}, Martin G. and {D'Odorico}, Valentina and {Bischetti}, Manuela and {Eilers}, Anna-Christina and {Keating}, Laura C. and {Kulkarni}, Girish and {Lai}, Samuel and {Mazzucchelli}, Chiara and {Qin}, Yuxiang and {Satyavolu}, Sindhu and {Wang}, Feige and {Yang}, Jinyi and {Zhu}, Yongda},
        title = "{Constraints on the Evolution of the Ionizing Background and Ionizing Photon Mean Free Path at the End of Reionization}",
      journal = {\apj},
         year = 2024,
        month = apr,
       volume = {965},
       number = {2},
          eid = {134},
        pages = {134},
          doi = {10.3847/1538-4357/ad1d5d},
archivePrefix = {arXiv},
       eprint = {2312.08464},
 primaryClass = {astro-ph.CO},
       adsurl = {https://ui.adsabs.harvard.edu/abs/2024ApJ...965..134D}
}

@ARTICLE{2023ApJ...955..115Z,
       author = {{Zhu}, Yongda and {Becker}, George D. and {Christenson}, Holly M. and {D'Aloisio}, Anson and {Bosman}, Sarah E.~I. and {Bakx}, Tom and {D'Odorico}, Valentina and {Bischetti}, Manuela and {Cain}, Christopher and {Davies}, Frederick B. and {Davies}, Rebecca L. and {Eilers}, Anna-Christina and {Fan}, Xiaohui and {Gaikwad}, Prakash and {Haehnelt}, Martin G. and {Keating}, Laura C. and {Kulkarni}, Girish and {Lai}, Samuel and {Ma}, Hai-Xia and {Mesinger}, Andrei and {Qin}, Yuxiang and {Satyavolu}, Sindhu and {Takeuchi}, Tsutomu T. and {Umehata}, Hideki and {Yang}, Jinyi},
        title = "{Probing Ultralate Reionization: Direct Measurements of the Mean Free Path over 5 $<$ z $<$ 6}",
      journal = {\apj},
         year = 2023,
        month = oct,
       volume = {955},
       number = {2},
          eid = {115},
        pages = {115},
          doi = {10.3847/1538-4357/aceef4},
archivePrefix = {arXiv},
       eprint = {2308.04614},
 primaryClass = {astro-ph.CO},
       adsurl = {https://ui.adsabs.harvard.edu/abs/2023ApJ...955..115Z}
}

@ARTICLE{2011ApJ...743...82M,
       author = {{McQuinn}, Matthew and {Oh}, S. Peng and {Faucher-Gigu{\`e}re}, Claude-Andr{\'e}},
        title = "{On Lyman-limit Systems and the Evolution of the Intergalactic Ionizing Background}",
      journal = {\apj},
         year = 2011,
        month = dec,
       volume = {743},
       number = {1},
          eid = {82},
        pages = {82},
          doi = {10.1088/0004-637X/743/1/82},
archivePrefix = {arXiv},
       eprint = {1101.1964},
 primaryClass = {astro-ph.CO},
       adsurl = {https://ui.adsabs.harvard.edu/abs/2011ApJ...743...82M}
}

@ARTICLE{2023MNRAS.525.4093G,
       author = {{Gaikwad}, Prakash and {Haehnelt}, Martin G. and {Davies}, Fredrick B. and {Bosman}, Sarah E.~I. and {Molaro}, Margherita and {Kulkarni}, Girish and {D'Odorico}, Valentina and {Becker}, George D. and {Davies}, Rebecca L. and {Nasir}, Fahad and {Bolton}, James S. and {Keating}, Laura C. and {Ir{\v{s}}i{\v{c}}}, Vid and {Puchwein}, Ewald and {Zhu}, Yongda and {Asthana}, Shikhar and {Yang}, Jinyi and {Lai}, Samuel and {Eilers}, Anna-Christina},
        title = "{Measuring the photoionization rate, neutral fraction, and mean free path of H I ionizing photons at 4.9 {\ensuremath{\leq}} z {\ensuremath{\leq}} 6.0 from a large sample of XShooter and ESI spectra}",
      journal = {\mnras},
         year = 2023,
        month = nov,
       volume = {525},
       number = {3},
        pages = {4093-4120},
          doi = {10.1093/mnras/stad2566},
archivePrefix = {arXiv},
       eprint = {2304.02038},
 primaryClass = {astro-ph.CO},
       adsurl = {https://ui.adsabs.harvard.edu/abs/2023MNRAS.525.4093G}
}

@ARTICLE{2021MNRAS.508.1853B,
       author = {{Becker}, George D. and {D'Aloisio}, Anson and {Christenson}, Holly M. and {Zhu}, Yongda and {Worseck}, G{\'a}bor and {Bolton}, James S.},
        title = "{The mean free path of ionizing photons at $5 < z < 6$: evidence for rapid evolution near reionization}",
      journal = {\mnras},
         year = 2021,
        month = dec,
       volume = {508},
       number = {2},
        pages = {1853-1869},
          doi = {10.1093/mnras/stab2696},
archivePrefix = {arXiv},
       eprint = {2103.16610},
 primaryClass = {astro-ph.CO},
       adsurl = {https://ui.adsabs.harvard.edu/abs/2021MNRAS.508.1853B}
}

@ARTICLE{2021ApJ...923..223Z,
       author = {{Zhu}, Yongda and {Becker}, George D. and {Bosman}, Sarah E.~I. and {Keating}, Laura C. and {Christenson}, Holly M. and {Ba{\~n}ados}, Eduardo and {Bian}, Fuyan and {Davies}, Frederick B. and {D'Odorico}, Valentina and {Eilers}, Anna-Christina and {Fan}, Xiaohui and {Haehnelt}, Martin G. and {Kulkarni}, Girish and {Pallottini}, Andrea and {Qin}, Yuxiang and {Wang}, Feige and {Yang}, Jinyi},
        title = "{Chasing the Tail of Cosmic Reionization with Dark Gap Statistics in the Ly{\ensuremath{\alpha}} Forest over 5 $<$ z $<$ 6}",
      journal = {\apj},
         year = 2021,
        month = dec,
       volume = {923},
       number = {2},
          eid = {223},
        pages = {223},
          doi = {10.3847/1538-4357/ac26c2},
archivePrefix = {arXiv},
       eprint = {2109.06295},
 primaryClass = {astro-ph.CO},
       adsurl = {https://ui.adsabs.harvard.edu/abs/2021ApJ...923..223Z}
}

@ARTICLE{2022ApJ...932...76Z,
       author = {{Zhu}, Yongda and {Becker}, George D. and {Bosman}, Sarah E.~I. and {Keating}, Laura C. and {D'Odorico}, Valentina and {Davies}, Rebecca L. and {Christenson}, Holly M. and {Ba{\~n}ados}, Eduardo and {Bian}, Fuyan and {Bischetti}, Manuela and {Chen}, Huanqing and {Davies}, Frederick B. and {Eilers}, Anna-Christina and {Fan}, Xiaohui and {Gaikwad}, Prakash and {Greig}, Bradley and {Haehnelt}, Martin G. and {Kulkarni}, Girish and {Lai}, Samuel and {Pallottini}, Andrea and {Qin}, Yuxiang and {Ryan-Weber}, Emma V. and {Walter}, Fabian and {Wang}, Feige and {Yang}, Jinyi},
        title = "{Long Dark Gaps in the Ly{\ensuremath{\beta}} Forest at z $<$ 6: Evidence of Ultra-late Reionization from XQR-30 Spectra}",
      journal = {\apj},
         year = 2022,
        month = jun,
       volume = {932},
       number = {2},
          eid = {76},
        pages = {76},
          doi = {10.3847/1538-4357/ac6e60},
archivePrefix = {arXiv},
       eprint = {2205.04569},
 primaryClass = {astro-ph.CO},
       adsurl = {https://ui.adsabs.harvard.edu/abs/2022ApJ...932...76Z}
}

@ARTICLE{2020MNRAS.491.1736K,
       author = {{Keating}, Laura C. and {Weinberger}, Lewis H. and {Kulkarni}, Girish and {Haehnelt}, Martin G. and {Chardin}, Jonathan and {Aubert}, Dominique},
        title = "{Long troughs in the Lyman-{\ensuremath{\alpha}} forest below redshift 6 due to islands of neutral hydrogen}",
      journal = {\mnras},
         year = 2020,
        month = jan,
       volume = {491},
       number = {2},
        pages = {1736-1745},
          doi = {10.1093/mnras/stz3083},
archivePrefix = {arXiv},
       eprint = {1905.12640},
 primaryClass = {astro-ph.CO},
       adsurl = {https://ui.adsabs.harvard.edu/abs/2020MNRAS.491.1736K}
}

@ARTICLE{2019MNRAS.485L..24K,
       author = {{Kulkarni}, Girish and {Keating}, Laura C. and {Haehnelt}, Martin G. and {Bosman}, Sarah E.~I. and {Puchwein}, Ewald and {Chardin}, Jonathan and {Aubert}, Dominique},
        title = "{Large Ly {\ensuremath{\alpha}} opacity fluctuations and low CMB {\ensuremath{\tau}} in models of late reionization with large islands of neutral hydrogen extending to z $<$ 5.5}",
      journal = {\mnras},
         year = 2019,
        month = may,
       volume = {485},
       number = {1},
        pages = {L24-L28},
          doi = {10.1093/mnrasl/slz025},
archivePrefix = {arXiv},
       eprint = {1809.06374},
 primaryClass = {astro-ph.CO},
       adsurl = {https://ui.adsabs.harvard.edu/abs/2019MNRAS.485L..24K}
}

@ARTICLE{2020MNRAS.494.3080N,
       author = {{Nasir}, Fahad and {D'Aloisio}, Anson},
        title = "{Observing the tail of reionization: neutral islands in the z = 5.5 Lyman-{\ensuremath{\alpha}} forest}",
      journal = {\mnras},
         year = 2020,
        month = may,
       volume = {494},
       number = {3},
        pages = {3080-3094},
          doi = {10.1093/mnras/staa894},
archivePrefix = {arXiv},
       eprint = {1910.03570},
 primaryClass = {astro-ph.CO},
       adsurl = {https://ui.adsabs.harvard.edu/abs/2020MNRAS.494.3080N}
}

@ARTICLE{2026arXiv260325788C,
  title     = "Introducing saguaro — simulating {IGM} evolution and environments
               at high resolution: setup and first results",
  author    = "Cain, Christopher and Das, Aloha and D'Aloisio, Anson and
               Foreman, Simon and Scannapieco, Evan and Moreno, Esteban and
               Lugatiman, Matthew and Cohon, Joshua and Tohfa, Hurum Maksora and
               Trac, Hy",
  journal   = "J. Cosmol. Astropart. Phys.",
  publisher = "IOP Publishing",
  volume    =  2026,
  number    =  09,
  pages     =  024,
  month     =  sep,
  year      =  2026
}

@ARTICLE{DOdorico2023,
       author = {{D'Odorico}, Valentina and {Ba{\~n}ados}, Eduardo and
                 {Becker}, George D. and et al.},
        title = "{XQR-30: the ultimate XSHOOTER quasar sample at the
                  reionization epoch}",
      journal = {\mnras},
         year = 2023,
        month = jul,
       volume = {523},
       number = {1},
        pages = {1399-1420},
          doi = {10.1093/mnras/stad1468},
archivePrefix = {arXiv},
       eprint = {2305.05053}
}

@ARTICLE{Bosman2021,
       author = {{Bosman}, Sarah E.~I. and {Davies}, Frederick B. and {Becker}, George D. and {Keating}, Laura C. and {Davies}, Rebecca L. and {Zhu}, Yongda and {Eilers}, Anna-Christina and {D'Odorico}, Valentina and {Bian}, Fuyan and {Bischetti}, Manuela and {Cristiani}, Stefano V. and {Fan}, Xiaohui and {Farina}, Emanuele P. and {Haehnelt}, Martin G. and {Hennawi}, Joseph F. and {Kulkarni}, Girish and {Mesinger}, Andrei and {Meyer}, Romain A. and {Onoue}, Masafusa and {Pallottini}, Andrea and {Qin}, Yuxiang and {Ryan-Weber}, Emma and {Schindler}, Jan-Torge and {Walter}, Fabian and {Wang}, Feige and {Yang}, Jinyi},
        title = "{Hydrogen reionization ends by z = 5.3: Lyman-{\ensuremath{\alpha}} optical depth measured by the XQR-30 sample}",
      journal = {\mnras},
         year = 2022,
        month = jul,
       volume = {514},
       number = {1},
        pages = {55-76},
          doi = {10.1093/mnras/stac1046},
archivePrefix = {arXiv},
       eprint = {2108.03699},
 primaryClass = {astro-ph.CO},
       adsurl = {https://ui.adsabs.harvard.edu/abs/2022MNRAS.514...55B}
}

@ARTICLE{Cain2021-yt,
  title     = "A Short Mean Free Path at z = 6 Favors Late and Rapid
               Reionization by Faint Galaxies",
  author    = "Cain, Christopher and D'Aloisio, Anson and Gangolli, Nakul and
               Becker, George D",
  journal   = "ApJL",
  publisher = "IOP Publishing",
  volume    =  917,
  number    =  2,
  pages     = "L37",
  month     =  aug,
  year      =  2021,
  language  = "en"
}

@ARTICLE{Chen2022,
  title     = "Patchy kinetic {Sunyaev–Zel’dovich} effect with controlled
               reionization history and morphology",
  author    = "Chen, Nianyi and Trac, Hy and Mukherjee, Suvodip and Cen, Renyue",
  journal   = "Astrophys. J.",
  publisher = "American Astronomical Society",
  volume    =  943,
  number    =  2,
  pages     =  138,
  month     =  feb,
  year      =  2023
}

@ARTICLE{DAloisio2018,
       author = {{D'Aloisio}, Anson and {McQuinn}, Matthew and {Davies}, Frederick B. and {Furlanetto}, Steven R.},
        title = "{Large fluctuations in the high-redshift metagalactic ionizing background}",
      journal = {\mnras},
         year = 2018,
        month = jan,
       volume = {473},
       number = {1},
        pages = {560-575},
          doi = {10.1093/mnras/stx2341},
archivePrefix = {arXiv},
       eprint = {1611.02711},
 primaryClass = {astro-ph.CO},
       adsurl = {https://ui.adsabs.harvard.edu/abs/2018MNRAS.473..560D}
}

@ARTICLE{Planck18,
       author = {{Planck Collaboration} and {Aghanim}, N. and {Akrami}, Y. and
         {Ashdown}, M. and {Aumont}, J. and {Baccigalupi}, C. and
         {Ballardini}, M. and {Banday}, A.~J. and {Barreiro}, R.~B. and
         {Bartolo}, N. and {Basak}, S. and {Battye}, R. and {Benabed}, K. and
         {Bernard}, J. -P. and {Bersanelli}, M. and {Bielewicz}, P. and
         {Bock}, J.~J. and {Bond}, J.~R. and {Borrill}, J. and {Bouchet}, F.~R. and
         {Boulanger}, F. and {Bucher}, M. and {Burigana}, C. and
         {Butler}, R.~C. and {Calabrese}, E. and {Cardoso}, J. -F. and
         {Carron}, J. and {Challinor}, A. and {Chiang}, H.~C. and {Chluba}, J. and
         {Colombo}, L.~P.~L. and {Combet}, C. and {Contreras}, D. and
         {Crill}, B.~P. and {Cuttaia}, F. and {de Bernardis}, P. and
         {de Zotti}, G. and {Delabrouille}, J. and {Delouis}, J. -M. and
         {Di Valentino}, E. and {Diego}, J.~M. and {Dor{\'e}}, O. and
         {Douspis}, M. and {Ducout}, A. and {Dupac}, X. and {Dusini}, S. and
         {Efstathiou}, G. and {Elsner}, F. and {En{\ss}lin}, T.~A. and
         {Eriksen}, H.~K. and {Fantaye}, Y. and {Farhang}, M. and
         {Fergusson}, J. and {Fernandez-Cobos}, R. and {Finelli}, F. and
         {Forastieri}, F. and {Frailis}, M. and {Fraisse}, A.~A. and
         {Franceschi}, E. and {Frolov}, A. and {Galeotta}, S. and {Galli}, S. and
         {Ganga}, K. and {G{\'e}nova-Santos}, R.~T. and {Gerbino}, M. and
         {Ghosh}, T. and {Gonz{\'a}lez-Nuevo}, J. and {G{\'o}rski}, K.~M. and
         {Gratton}, S. and {Gruppuso}, A. and {Gudmundsson}, J.~E. and
         {Hamann}, J. and {Handley}, W. and {Hansen}, F.~K. and {Herranz}, D. and
         {Hildebrandt}, S.~R. and {Hivon}, E. and {Huang}, Z. and
         {Jaffe}, A.~H. and {Jones}, W.~C. and {Karakci}, A. and
         {Keih{\"a}nen}, E. and {Keskitalo}, R. and {Kiiveri}, K. and {Kim}, J. and
         {Kisner}, T.~S. and {Knox}, L. and {Krachmalnicoff}, N. and {Kunz}, M. and
         {Kurki-Suonio}, H. and {Lagache}, G. and {Lamarre}, J. -M. and
         {Lasenby}, A. and {Lattanzi}, M. and {Lawrence}, C.~R. and
         {Le Jeune}, M. and {Lemos}, P. and {Lesgourgues}, J. and {Levrier}, F. and
         {Lewis}, A. and {Liguori}, M. and {Lilje}, P.~B. and {Lilley}, M. and
         {Lindholm}, V. and {L{\'o}pez-Caniego}, M. and {Lubin}, P.~M. and
         {Ma}, Y. -Z. and {Mac{\'\i}as-P{\'e}rez}, J.~F. and {Maggio}, G. and
         {Maino}, D. and {Mandolesi}, N. and {Mangilli}, A. and
         {Marcos-Caballero}, A. and {Maris}, M. and {Martin}, P.~G. and
         {Martinelli}, M. and {Mart{\'\i}nez-Gonz{\'a}lez}, E. and
         {Matarrese}, S. and {Mauri}, N. and {McEwen}, J.~D. and
         {Meinhold}, P.~R. and {Melchiorri}, A. and {Mennella}, A. and
         {Migliaccio}, M. and {Millea}, M. and {Mitra}, S. and
         {Miville-Desch{\^e}nes}, M. -A. and {Molinari}, D. and {Montier}, L. and
         {Morgante}, G. and {Moss}, A. and {Natoli}, P. and
         {N{\o}rgaard-Nielsen}, H.~U. and {Pagano}, L. and {Paoletti}, D. and
         {Partridge}, B. and {Patanchon}, G. and {Peiris}, H.~V. and
         {Perrotta}, F. and {Pettorino}, V. and {Piacentini}, F. and
         {Polastri}, L. and {Polenta}, G. and {Puget}, J. -L. and
         {Rachen}, J.~P. and {Reinecke}, M. and {Remazeilles}, M. and
         {Renzi}, A. and {Rocha}, G. and {Rosset}, C. and {Roudier}, G. and
         {Rubi{\~n}o-Mart{\'\i}n}, J.~A. and {Ruiz-Granados}, B. and
         {Salvati}, L. and {Sandri}, M. and {Savelainen}, M. and {Scott}, D. and
         {Shellard}, E.~P.~S. and {Sirignano}, C. and {Sirri}, G. and
         {Spencer}, L.~D. and {Sunyaev}, R. and {Suur-Uski}, A. -S. and
         {Tauber}, J.~A. and {Tavagnacco}, D. and {Tenti}, M. and
         {Toffolatti}, L. and {Tomasi}, M. and {Trombetti}, T. and
         {Valenziano}, L. and {Valiviita}, J. and {Van Tent}, B. and
         {Vibert}, L. and {Vielva}, P. and {Villa}, F. and {Vittorio}, N. and {Wand
        elt}, B.~D. and {Wehus}, I.~K. and {White}, M. and {White}, S.~D.~M. and
         {Zacchei}, A. and {Zonca}, A.},
        title = "{Planck 2018 results. VI. Cosmological parameters}",
      journal = {\aap},
         year = 2020,
        month = sep,
       volume = {641},
          eid = {A6},
        pages = {A6},
          doi = {10.1051/0004-6361/201833910},
archivePrefix = {arXiv},
       eprint = {1807.06209},
 primaryClass = {astro-ph.CO},
       adsurl = {https://ui.adsabs.harvard.edu/abs/2020A&A...641A...6P}
}

@ARTICLE{Kramer2025,
  title     = "Cross-correlating the patchy screening and kinetic
               {Sunyaev–Zel’dovich} effects as a new probe of reionization",
  author    = "Kramer, Darby and van Engelen, Alexander and Cain, Christopher
               and MacCrann, Niall and Trac, Hy and Grayson, Skylar and
               Scannapieco, Evan and Sherwin, Blake",
  journal   = "Astrophys. J.",
  publisher = "American Astronomical Society",
  volume    =  994,
  number    =  1,
  pages     =  82,
  month     =  nov,
  year      =  2025,
  language  = "en"
}

@ARTICLE{Cain2024-wt,
  title     = "Chasing the beginning of reionization in the {JWST} era",
  author    = "Cain, Christopher and Lopez, Garett and D'Aloisio, Anson and
               Muñoz, Julian B and Jansen, Rolf A and Windhorst, Rogier A and
               Gangolli, Nakul",
  journal   = "Astrophys. J.",
  publisher = "American Astronomical Society",
  volume    =  980,
  number    =  1,
  pages     =  83,
  month     =  feb,
  year      =  2025,
  language  = "en"
}

@ARTICLE{McGreer2011-zm,
  title     = "The first (nearly) model-independent constraint on the neutral
               hydrogen fraction at: Upper limit on the neutral fraction at z $\sim$
               5-6",
  author    = "McGreer, Ian D and Mesinger, Andrei and Fan, Xiaohui",
  journal   = "Mon. Not. R. Astron. Soc.",
  publisher = "Oxford University Press (OUP)",
  volume    =  415,
  number    =  4,
  pages     = "3237--3246",
  month     =  aug,
  year      =  2011,
  language  = "en"
}

@ARTICLE{2022JCAP...10..007S,
       author = {{Sailer}, Noah and {Chen}, Shi-Fan and {White}, Martin},
        title = "{Optical depth to reionization from perturbative 21 cm clustering}",
      journal = {\jcap},
         year = 2022,
        month = oct,
       volume = {2022},
       number = {10},
          eid = {007},
        pages = {007},
          doi = {10.1088/1475-7516/2022/10/007},
archivePrefix = {arXiv},
       eprint = {2205.11504},
 primaryClass = {astro-ph.CO},
       adsurl = {https://ui.adsabs.harvard.edu/abs/2022JCAP...10..007S}
}

@ARTICLE{Trac2021-xm,
  title     = "{AMBER}: A semi-numerical abundance matching box for the Epoch of
               reionization",
  author    = "Trac, Hy and Chen, Nianyi and Holst, Ian and Alvarez, Marcelo A
               and Cen, Renyue",
  journal   = "Astrophys. J.",
  publisher = "American Astronomical Society",
  volume    =  927,
  number    =  2,
  pages     =  186,
  month     =  mar,
  year      =  2022,
  language  = "en"
}

@ARTICLE{Zhu2024-aq,
  title     = "Damping wing-like features in the stacked Ly $\alpha$ forest: Potential
               neutral hydrogen islands at \textit{z} $<$ 6",
  author    = "Zhu, Yongda and Becker, George D and Bosman, Sarah E I and Cain,
               Christopher and Keating, Laura C and Nasir, Fahad and D'Odorico,
               Valentina and Bañados, Eduardo and Bian, Fuyan and Bischetti,
               Manuela and Bolton, James S and Chen, Huanqing and D'Aloisio,
               Anson and Davies, Frederick B and Davies, Rebecca L and Eilers,
               Anna-Christina and Fan, Xiaohui and Gaikwad, Prakash and Greig,
               Bradley and Haehnelt, Martin G and Kulkarni, Girish and Lai,
               Samuel and Puchwein, Ewald and Qin, Yuxiang and Ryan-Weber, Emma
               V and Satyavolu, Sindhu and Spina, Benedetta and Walter, Fabian
               and Wang, Feige and Wolfson, Molly and Yang, Jinyi",
  journal   = "Mon. Not. R. Astron. Soc. Lett.",
  publisher = "Oxford University Press (OUP)",
  volume    =  533,
  number    =  1,
  pages     = "L49--L56",
  month     =  jun,
  year      =  2024,
  language  = "en"
}

@ARTICLE{2020ApJ...898..149D,
       author = {{D'Aloisio}, Anson and {McQuinn}, Matthew and {Trac}, Hy and {Cain}, Christopher and {Mesinger}, Andrei},
        title = "{Hydrodynamic Response of the Intergalactic Medium to Reionization}",
      journal = {\apj},
         year = 2020,
        month = aug,
       volume = {898},
       number = {2},
          eid = {149},
        pages = {149},
          doi = {10.3847/1538-4357/ab9f2f},
archivePrefix = {arXiv},
       eprint = {2002.02467},
 primaryClass = {astro-ph.CO},
       adsurl = {https://ui.adsabs.harvard.edu/abs/2020ApJ...898..149D}
}

@ARTICLE{2021ApJ...923..161N,
       author = {{Nasir}, Fahad and {Cain}, Christopher and {D'Aloisio}, Anson and {Gangolli}, Nakul and {McQuinn}, Matthew},
        title = "{Hydrodynamic Response of the Intergalactic Medium to Reionization. II. Physical Characteristics and Dynamics of Ionizing Photon Sinks}",
      journal = {\apj},
         year = 2021,
        month = dec,
       volume = {923},
       number = {2},
          eid = {161},
        pages = {161},
          doi = {10.3847/1538-4357/ac2eb9},
archivePrefix = {arXiv},
       eprint = {2108.04837},
 primaryClass = {astro-ph.CO},
       adsurl = {https://ui.adsabs.harvard.edu/abs/2021ApJ...923..161N}
}

@ARTICLE{Gaikwad2025,
       author = {{Gaikwad}, Prakash and {Davies}, Fredrick B. and {Haehnelt}, Martin G.},
        title = "{Measuring photo-ionization rate and mean free path of He II ionizing photons at 2.5 {\ensuremath{\leq}} z {\ensuremath{\leq}} 3.6: evidence for late and rapid HeII reionization part II}",
      journal = {\jcap},
         year = 2025,
        month = sep,
       volume = {2025},
       number = {9},
          eid = {057},
        pages = {057},
          doi = {10.1088/1475-7516/2025/09/057},
archivePrefix = {arXiv},
       eprint = {2503.04893},
 primaryClass = {astro-ph.CO},
       adsurl = {https://ui.adsabs.harvard.edu/abs/2025JCAP...09..057G}
}

@ARTICLE{2016MNRAS.460.1885U,
       author = {{Upton Sanderbeck}, Phoebe R. and {D'Aloisio}, Anson and {McQuinn}, Matthew J.},
        title = "{Models of the thermal evolution of the intergalactic medium after reionization}",
      journal = {\mnras},
         year = 2016,
        month = aug,
       volume = {460},
       number = {2},
        pages = {1885-1897},
          doi = {10.1093/mnras/stw1117},
archivePrefix = {arXiv},
       eprint = {1511.05992},
 primaryClass = {astro-ph.CO},
       adsurl = {https://ui.adsabs.harvard.edu/abs/2016MNRAS.460.1885U}
}

@ARTICLE{Davies2025-lf,
  title     = "Updated dark pixel fraction constraints on reionization’s end
               from the Lyman-series forests of {XQR-30}",
  author    = "Davies, Frederick B and Bosman, Sarah E I and D'Odorico,
               Valentina and Campo, Sofia and Mesinger, Andrei and Qin, Yuxiang
               and Becker, George D and Bañados, Eduardo and Chen, Huanqing and
               Cristiani, Stefano and Fan, Xiaohui and Gallerani, Simona and
               Haehnelt, Martin G and Keating, Laura C and Lai, Samuel and
               Ryan-Weber, Emma and Wang, Feige and Yang, Jinyi and Zhu, Yongda",
  journal   = "Mon. Not. R. Astron. Soc.",
  publisher = "Oxford University Press (OUP)",
  volume    =  545,
  number    =  2,
  pages     = "staf1862",
  month     =  dec,
  year      =  2025,
  language  = "en"
}

@ARTICLE{Millea2018-xl,
  title     = "Cosmic microwave background constraints in light of priors over
               reionization histories",
  author    = "Millea, Marius and Bouchet, François",
  journal   = "Astron. Astrophys.",
  publisher = "EDP Sciences",
  volume    =  617,
  pages     = "A96",
  month     =  sep,
  year      =  2018,
  language  = "en"
}

@ARTICLE{Qin2020-kf,
  title     = "Reionization inference from the {CMB} optical depth and {E}-mode
               polarization power spectra",
  author    = "Qin, Yuxiang and Poulin, Vivian and Mesinger, Andrei and Greig,
               Bradley and Murray, Steven and Park, Jaehong",
  journal   = "Mon. Not. R. Astron. Soc.",
  publisher = "Oxford University Press (OUP)",
  volume    =  499,
  number    =  1,
  pages     = "550--558",
  month     =  oct,
  year      =  2020,
  language  = "en"
}

@ARTICLE{Heinrich2021-ik,
  title     = "Reionization effective likelihood from \textit{Planck} 2018 data",
  author    = "Heinrich, Chen and Hu, Wayne",
  journal   = "Phys. Rev. D.",
  publisher = "American Physical Society (APS)",
  volume    =  104,
  number    =  6,
  pages     =  063505,
  month     =  sep,
  year      =  2021,
  language  = "en"
}

@ARTICLE{Wu2021-od,
  title     = "The high-redshift tail of stellar reionization in {LCDM} is
               beyond the reach of the low-$\ell$ {CMB}",
  author    = "Wu, Xiaohan and McQuinn, Matthew and Eisenstein, Daniel and
               Iršič, Vid",
  journal   = "Mon. Not. R. Astron. Soc.",
  publisher = "Oxford University Press (OUP)",
  volume    =  508,
  number    =  2,
  pages     = "2784--2797",
  month     =  oct,
  year      =  2021,
  language  = "en"
}

@ARTICLE{Ilic2025-xj,
  title     = "Reconstructing the epoch of reionisation with Planck {PR4}",
  author    = "Ilić, S and Tristram, M and Douspis, M and Gorce, A and
               Henrot-Versillé, S and Hergt, L T and Langer, M and McBride, L
               and Muñoz-Echeverría, M and Pointecouteau, E and Salvati, L",
  journal   = "Astron. Astrophys.",
  publisher = "EDP Sciences",
  volume    =  700,
  pages     = "A26",
  month     =  jul,
  year      =  2025,
  language  = "en"
}

@ARTICLE{Jin2023-bh,
  title     = "(Nearly) Model-independent constraints on the neutral hydrogen
               fraction in the intergalactic medium at z $\sim$ {5–7} using dark
               pixel fractions in ly$\alpha$ and Ly$\beta$ forests",
  author    = "Jin, Xiangyu and Yang, Jinyi and Fan, Xiaohui and Wang, Feige and
               Bañados, Eduardo and Bian, Fuyan and Davies, Frederick B and
               Eilers, Anna-Christina and Farina, Emanuele Paolo and Hennawi,
               Joseph F and Pacucci, Fabio and Venemans, Bram and Walter, Fabian",
  journal   = "Astrophys. J.",
  publisher = "American Astronomical Society",
  volume    =  942,
  number    =  2,
  pages     =  59,
  month     =  jan,
  year      =  2023,
  language  = "en"
}

@ARTICLE{Malloy2015-ko,
  title     = "{HOW} {TO} {SEARCH} {FOR} {ISLANDS} {OF} {NEUTRAL} {HYDROGEN}
               {IN} {THE z$\sim$} 5.5 {IGM}",
  author    = "Malloy, Matthew and Lidz, Adam",
  journal   = "Astrophys. J.",
  publisher = "IOP Publishing",
  volume    =  799,
  number    =  2,
  pages     =  179,
  month     =  jan,
  year      =  2015,
  language  = "en"
}

@ARTICLE{Gaikwad2023-mm,
  title     = "Measuring the photo-ionization rate, neutral fraction and mean
               free path of {H} i ionizing photons at 4.9 $\leq$ \textit{z} $\leq$ 6.0
               from a large sample of {XShooter} and {ESI} spectra",
  author    = "Gaikwad, Prakash and Haehnelt, Martin G and Davies, Fredrick B
               and Bosman, Sarah E I and Molaro, Margherita and Kulkarni, Girish
               and D'Odorico, Valentina and Becker, George D and Davies, Rebecca
               L and Nasir, Fahad and Bolton, James S and Keating, Laura C and
               Iršič, Vid and Puchwein, Ewald and Zhu, Yongda and Asthana,
               Shikhar and Yang, Jinyi and Lai, Samuel and Eilers,
               Anna-Christina",
  journal   = "Mon. Not. R. Astron. Soc.",
  publisher = "Oxford University Press (OUP)",
  volume    =  525,
  number    =  3,
  pages     = "4093--4120",
  month     =  sep,
  year      =  2023,
  language  = "en"
}

@ARTICLE{Cain2026-op,
  title         = "A self-consistent analytical model for both the
                   photoionization rate and reionization history",
  author        = "Cain, Christopher and Croker, Kevin S and D'Aloisio, Anson
                   and Georgiev, Ivelin and Tohfa, Hurum Maksora and Zhu, Yongda
                   and Windhorst, Rogier",
  journal       = "arXiv [astro-ph.CO]",
  month         =  jun,
  year          =  2026,
  archivePrefix = "arXiv",
  primaryClass  = "astro-ph.CO",
  language      = "en"
}

@article{Lewis:2019xzd,
   author = "Lewis, Antony",
   title = "{GetDist: a Python package for analysing Monte Carlo samples}",
   eprint = "1910.13970",
   archivePrefix = "arXiv",
   primaryClass = "astro-ph.IM",
   doi = "10.1088/1475-7516/2025/08/025",
   journal = "JCAP",
   volume = "08",
   pages = "025",
   year = "2025"
}

@ARTICLE{Chaubal2026-wu,
  title         = "{SPT}-{3G} {D1}: A Measurement of Secondary Cosmic Microwave
                   Background Anisotropy Power",
  author        = "Chaubal, P and Huang, N and Reichardt, C L and Anderson, A J
                   and Ansarinejad, B and Archipley, M and Balkenhol, L and
                   Barron, D R and Benabed, K and Bender, A N and Benson, B A
                   and Bianchini, F and Bleem, L E and Bocquet, S and Bouchet, F
                   R and Bryant, L and Camphuis, E and Campitiello, M G and
                   Carlstrom, J E and Carron, J and Chang, C L and Chichura, P M
                   and Chokshi, A and Chou, T-L and Coerver, A and Crawford, T M
                   and Daley, C and de Haan, T and Dibert, K R and Dobbs, M A
                   and Doohan, M and Doussot, A and Dutcher, D and Everett, W
                   and Feng, C and Ferguson, K R and Ferree, N C and Fichman, K
                   and Foster, A and Galli, S and Gambrel, A E and Gao, A K and
                   Gardner, R W and Ge, F and Goeckner-Wald, N and Gualtieri, R
                   and Guidi, F and Guns, S and Halverson, N W and Hivon, E and
                   Ho, A Y Q and Holder, G P and Holzapfel, W L and Hood, J C
                   and Hryciuk, A and Jhaveri, T and Kéruzoré, F and Khalife, A
                   R and Knox, L and Korman, M and Kornoelje, K and Kuo, C-L and
                   Levy, K and Li, Y and Lowitz, A E and Lu, C and Lynch, G P
                   and Maccarone, T J and Maniyar, A S and Martsen, E S and
                   Menanteau, F and Millea, M and Montgomery, J and Nakato, Y
                   and Natoli, T and Noble, G I and Omori, Y and Ouellette, A
                   and Pan, Z and Paschos, P and Phadke, K A and Pollak, A W and
                   Prabhu, K and Quan, W and Raghunathan, S and Rahimi, M and
                   Rahlin, A and Rouble, M and Ruhl, J E and Schiappucci, E and
                   Silva Oliveira, A C and Simpson, A and Sobrin, J A and Stark,
                   A A and Stephen, J and Tandoi, C and Thorne, B and
                   Trendafilova, C and Umilta, C and Vieira, J D and Vieregg, A
                   G and Vitrier, A and Wan, Y and Whitehorn, N and Wu, W L K
                   and Young, M R and Zebrowski, J A",
  journal       = "arXiv [astro-ph.CO]",
  pages         = "arXiv:2601.20551",
  month         =  jan,
  year          =  2026,
  archivePrefix = "arXiv",
  primaryClass  = "astro-ph.CO"
}

@ARTICLE{2026arXiv260518939S,
  title         = "The Lumina project: {CMB} optical depth fluctuations from
                   patchy reionization",
  author        = "Smith, Aaron and Zier, Oliver and Kannan, Rahul and Shen,
                   Xuejian and Liu, Rongrong and Vogelsberger, Mark and
                   Springel, Volker and Pakmor, Ruediger and Koehler, Sonja M
                   and Hernquist, Lars and Neyer, Meredith",
  journal       = "arXiv [astro-ph.CO]",
  pages         = "arXiv:2605.18939",
  month         =  may,
  year          =  2026,
  archivePrefix = "arXiv",
  primaryClass  = "astro-ph.CO"
}

@ARTICLE{2018RNAAS...2..135G,
       author = {{Glazer}, Daniel and {Rau}, Markus Michael and {Trac}, Hy},
        title = "{The Reionization Parameter Space Consistent with the Thomson Optical Depth from Planck}",
      journal = {Research Notes of the American Astronomical Society},
         year = 2018,
        month = aug,
       volume = {2},
       number = {3},
          eid = {135},
        pages = {135},
          doi = {10.3847/2515-5172/aad68a},
archivePrefix = {arXiv},
       eprint = {1808.00553},
 primaryClass = {astro-ph.CO},
       adsurl = {https://ui.adsabs.harvard.edu/abs/2018RNAAS...2..135G}
}

@ARTICLE{1994ApJ...427...25S,
       author = {{Shapiro}, Paul R. and {Giroux}, Mark L. and {Babul}, Arif},
        title = "{Reionization in a Cold Dark Matter Universe: The Feedback of Galaxy Formation on the Intergalactic Medium}",
      journal = {\apj},
         year = 1994,
        month = may,
       volume = {427},
        pages = {25},
          doi = {10.1086/174120},
       adsurl = {https://ui.adsabs.harvard.edu/abs/1994ApJ...427...25S}
}

@ARTICLE{2022MNRAS.515.2386R,
       author = {{Rosdahl}, Joakim and {Blaizot}, J{\'e}r{\'e}my and {Katz}, Harley and {Kimm}, Taysun and {Garel}, Thibault and {Haehnelt}, Martin and {Keating}, Laura C. and {Martin-Alvarez}, Sergio and {Michel-Dansac}, L{\'e}o and {Ocvirk}, Pierre},
        title = "{LyC escape from SPHINX galaxies in the Epoch of Reionization}",
      journal = {\mnras},
         year = 2022,
        month = sep,
       volume = {515},
       number = {2},
        pages = {2386-2414},
          doi = {10.1093/mnras/stac1942},
archivePrefix = {arXiv},
       eprint = {2207.03232},
 primaryClass = {astro-ph.GA},
       adsurl = {https://ui.adsabs.harvard.edu/abs/2022MNRAS.515.2386R}
}

@ARTICLE{2022A&A...663A..66M,
       author = {{Maji}, Moupiya and {Verhamme}, Anne and {Rosdahl}, Joakim and {Garel}, Thibault and {Blaizot}, J{\'e}r{\'e}my and {Mauerhofer}, Valentin and {Pittavino}, Marta and {Victoria Feser}, Maria-Pia and {Chuniaud}, Mathieu and {Kimm}, Taysun and {Katz}, Harley and {Haehnelt}, Martin},
        title = "{Predicting Lyman-continuum emission of galaxies using their physical and Lyman-alpha emission properties}",
      journal = {\aap},
         year = 2022,
        month = jul,
       volume = {663},
          eid = {A66},
        pages = {A66},
          doi = {10.1051/0004-6361/202142740},
archivePrefix = {arXiv},
       eprint = {2204.02440},
 primaryClass = {astro-ph.GA},
       adsurl = {https://ui.adsabs.harvard.edu/abs/2022A&A...663A..66M}
}

@ARTICLE{2023MNRAS.518..270K,
       author = {{Katz}, Harley and {Saxena}, Aayush and {Rosdahl}, Joki and {Kimm}, Taysun and {Blaizot}, Jeremy and {Garel}, Thibault and {Michel-Dansac}, Leo and {Haehnelt}, Martin and {Ellis}, Richard S. and {Pentericci}, Laura and {Devriendt}, Julien and {Slyz}, Adrianne},
        title = "{Two modes of LyC escape from bursty star formation: implications for [C II] deficits and the sources of reionization}",
      journal = {\mnras},
         year = 2023,
        month = jan,
       volume = {518},
       number = {1},
        pages = {270-285},
          doi = {10.1093/mnras/stac3019},
archivePrefix = {arXiv},
       eprint = {2210.09156},
 primaryClass = {astro-ph.GA},
       adsurl = {https://ui.adsabs.harvard.edu/abs/2023MNRAS.518..270K}
}
\end{document}